\documentclass{article}

\usepackage{microtype}
\usepackage{graphicx}
\graphicspath{{images/}}
\usepackage{booktabs} 
\usepackage{cite}
\usepackage{mathtools}
\usepackage{amsmath,amssymb,amsfonts}
\usepackage{textcomp}
\usepackage{pstricks}
\usepackage{mathdots}
\usepackage{subcaption}
\usepackage{multirow}
\usepackage{epic}
\usepackage{eepic}
\usepackage{paralist}
\newcommand{\remove}[1]{}

\def\math#1{$#1$}

\def\mand#1{$$#1$$}
\def\mandc#1{\mand{\abovedisplayskip=3pt plus 1pt minus 1pt%
\abovedisplayshortskip=0pt plus 1pt minus 1pt%
\belowdisplayskip=3pt plus 1pt minus 1pt%
\belowdisplayshortskip=0pt plus 1pt minus 1pt%
#1}}

\def\frac#1#2{{#1\over #2}}

\def\mld#1{\begin{equation}
#1
\end{equation}}
\def\mldc#1{\mld{\abovedisplayskip=3pt plus 1pt minus 1pt%
\abovedisplayshortskip=0pt plus 1pt minus 1pt%
\belowdisplayskip=3pt plus 1pt minus 1pt%
\belowdisplayshortskip=0pt plus 1pt minus 1pt%
#1}}

\DeclareSymbolFont{AMSb}{U}{msb}{m}{n}
\DeclareMathSymbol{\N}{\mathbin}{AMSb}{"4E}
\DeclareMathSymbol{\Z}{\mathbin}{AMSb}{"5A}
\DeclareMathSymbol{\R}{\mathbin}{AMSb}{"52}
\DeclareMathSymbol{\Q}{\mathbin}{AMSb}{"51}
\DeclareMathSymbol{\I}{\mathbin}{AMSb}{"49}
\DeclareMathSymbol{\C}{\mathbin}{AMSb}{"43}

\def\Z{{\mathbf Z}}

\def\r#1{{[\ref{#1}]}}

\newtheorem{theorem}{\bf Theorem}[section]

\newtheorem{definition}[theorem]{Definition}

\newcounter{exercisenum}

\newenvironment{example}[1]{
\begin{quotation}
\noindent {\bf Example{#1}:}
}
{
\end{quotation}
\noindent}

\def\dotfil{\leaders\hbox to 1.5mm{.}\hfill}
\newcounter{rmnum}
\def\RN#1{\setcounter{rmnum}{#1}\uppercase\expandafter{\romannumeral\value{rmnum}}}
\def\rn#1{\setcounter{rmnum}{#1}\expandafter{\romannumeral\value{rmnum}}}

\usepackage{hyperref}

\usepackage[accepted]{icml2018}

\icmltitlerunning{The Intrinsic Scale Of Networks Is Small}

\begin{document}

\twocolumn[
\icmltitle{The Intrinsic Scale Of Networks Is Small}




\begin{icmlauthorlist}
\icmlauthor{Malik Magdon-Ismail}{rpi}
\icmlauthor{Kshiteesh Hegde}{rpi}
\end{icmlauthorlist}

\icmlaffiliation{rpi}{Department of Computer Science, Rensselaer Polytechnic Institute, Troy, NY, USA}

\icmlcorrespondingauthor{Malik Magdon-Ismail}{magdon@cs.rpi.edu}

\icmlkeywords{structure, signature, deep learning, convolutional neural networks}

\vskip 0.3in
]



\printAffiliationsAndNotice{} 

\begin{abstract}
We define the intrinsic scale 
at which a network begins to reveal its identity
as the scale at which subgraphs in the network (created by a random walk)
are distinguishable from similar sized subgraphs in
a perturbed copy of the network.
We conduct an extensive study of intrinsic scale for several
networks, ranging from structured (e.g. road networks) to
ad-hoc and unstructured (e.g. crowd sourced information networks),
to biological.
We find:
(a) The intrinsic scale is surprisingly small
(7-20 vertices), even though the networks are many orders of magnitude larger.
(b) The intrinsic scale quantifies ``structure'' in a
network -- networks which are explicitly constructed for
specific tasks have smaller intrinsic scale.
(c) The structure at different scales can be fragile (easy
to disrupt) or robust.

\end{abstract}

Large networks are ubiquitous,
either explicitly (e.g. the Facebook social network) or implicitly
(e.g. the DBLP citation data induces a network of researchers; the
Amazon purchase data induces a product network).
Significant effort has been spent quantifying a network's topological structure.
Seidman~\cite{seidman1983} computes network cohesion
using minimum vertex cuts.
Reagans et al \cite{reagans2003} view network structure as facilitating knowledge transfer and argue that social ties, cohesion and network range play important roles. Olbrich et al \cite{olbrich2010} use exponential families to estimate degree distribution, clustering and assortativity coefficients, and subgraph densities.
Clustering the vertices based on the topology is a powerful tool
for uncovering structure. Newman in \cite{newman2004,newman8577} developed a popular approach to non-overlapping clustering, which optimizes a
modularity objective that (globally) quantifies the quality of the entire collection of clusters. Some of the earliest work which allows
overlapping clusters is based on defining a cluster as a locally optimal set
(different locally optimal sets may overlap), \cite{malik62,malik47}. We refer to \cite{F2010} for a survey on
clustering.

The trend is to classify structure using global aggregate
parameters (e.g. power laws)
which emerge in the  large scale limit. We tackle the opposite end of
the spectrum, and ask:
\mandc{
\text{\emph{At what
(small) scale does a network identify itself?}}}
We propose a methodology which, given a network \math{N} with
\math{n} vertices and \math{m} edges, extracts
the intrinsic scale.
Results from several networks reveal
a surprising conclusion:
\mandc{\text{\emph{The intrinsic scale of real networks is
7-20 vertices.}}}
Networks have non-trivial structure at small-scales, where
aggregate  parameters such as power-law exponents
aren't stable.

\paragraph{Intrinsic Scale Via Distinguishability of Subgraphs
Induced by Random Walks.}
We argue that
a network \math{N} has structure at scale \math{\kappa} if
typical size-\math{\kappa} subgraphs
from \math{N} are distinguishable from size-\math{\kappa} subgraphs
in a randomized copy of \math{N}. This distinguishability
implies ``something'' in~\math{N} at scale \math{\kappa} must have been
disturbed.
Let \math{N_\delta} be a perturbed copy of \math{N}
with the same degree distribution, where \math{\delta} quantifies the extent
of the perturbation. In particular,
\math{N_0=N} and \math{N_\infty} is a random graph
with the same degrees as \math{N}.
To construct perturbed graphs
\math{N_\delta} for \math{\delta=1,2,\ldots}, we use 
\math{\delta} random edge-swaps to
rewire the network.
In an edge swap, edges \math{(u,v)} and \math{(x,y)} with
distinct vertices \math{u,v,x,y} are rewired as follows:
\mandc{
\begin{array}[t]{ccc}
(u,v)\quad(x,y)&\quad\rightarrow\quad&(u,x)\quad(v,y)\\
\begin{picture}(22,18)
\thicklines
\put(0,0){\circle{5}}
\put(0,12){\circle{5}}
\put(0,2.5){\line(0,1){7}}
\put(22,0){\circle*{5}}
\put(22,12){\circle*{5}}
\put(22,2.5){\line(0,1){7}}
\put(-2.5,0){\line(-3,-1){8}}
\put(-2.5,0){\line(-3,0){8}}
\put(-2.5,0){\line(-3,1){8}}
\put(-2.5,12){\line(-3,-1){8}}
\put(-2.5,12){\line(-3,1){8}}
\put(24.5,12){\line(3,-2){8}}
\put(24.5,12){\line(3,-1){8}}
\put(24.5,12){\line(3,0){8}}
\put(24.5,12){\line(3,1){8}}
\put(24.5,0){\line(3,-1){8}}
\put(24.5,0){\line(3,1){8}}
\end{picture}
&&
\begin{picture}(22,18)
\thicklines
\put(0,0){\circle{5}}
\put(0,12){\circle{5}}
\put(2.5,0){\line(1,0){17}}
\put(22,0){\circle*{5}}
\put(22,12){\circle*{5}}
\put(2.5,12){\line(1,0){17}}
\put(-2.5,0){\line(-3,-1){8}}
\put(-2.5,0){\line(-3,0){8}}
\put(-2.5,0){\line(-3,1){8}}
\put(-2.5,12){\line(-3,-1){8}}
\put(-2.5,12){\line(-3,1){8}}
\put(24.5,12){\line(3,-2){8}}
\put(24.5,12){\line(3,-1){8}}
\put(24.5,12){\line(3,0){8}}
\put(24.5,12){\line(3,1){8}}
\put(24.5,0){\line(3,-1){8}}
\put(24.5,0){\line(3,1){8}}
\end{picture}
\end{array},
}
Observe that an edge swap preserves every vertex-degree.
We illustrate a sequence of edge swaps on a toy graph below.

\centerline{\setlength{\unitlength}{1.1pt}\thicklines%
\begin{tabular}{ccccccccc}
\begin{picture}(16,30)(-6,-8)
\put(0,0){\circle*{3}}
\put(0,15){\circle*{3}}
\put(-6,8){\circle*{3}}
\put(-10,2){\circle*{3}}
\put(-4,-8){\circle*{3}}
\put(5,-8){\circle*{3}}
\put(10,2){\circle*{3}}
\put(5,8){\circle*{3}}
\put(0,0){\line(0,1){15}}
\put(0,0){\line(5,-8){5}}
\put(0,0){\line(-5,1){10}}
\put(-4,-8){\line(1,0){9}}
\put(-10,2){\line(6,-10){6}}
\put(5,-8){\line(1,2){5}}
\put(10,2){\line(-5,6){5}}
\put(0,15){\line(-6,-7){6}}
\end{picture}
&\math{\rightarrow}&
\begin{picture}(16,30)(-6,-8)
\put(0,0){\circle*{3}}
\put(0,15){\circle*{3}}
\put(-6,8){\circle*{3}}
\put(-10,2){\circle*{3}}
\put(-4,-8){\circle*{3}}
\put(5,-8){\circle*{3}}
\put(10,2){\circle*{3}}
\put(5,8){\circle*{3}}
\put(0,0){\line(0,1){15}}
\put(0,0){\line(5,-8){5}}
\put(0,0){\line(-5,1){10}}
\put(-4,-8){\line(1,0){9}}
\put(-10,2){\line(6,-10){6}}
\put(5,-8){\line(1,2){5}}
\put(0,15){\line(5,-7){5}}
\qbezier(-6,8)(2,35)(10,2)
\end{picture}
&\math{\rightarrow}&
\begin{picture}(16,30)(-6,-8)
\put(0,0){\circle*{3}}
\put(0,15){\circle*{3}}
\put(-6,8){\circle*{3}}
\put(-10,2){\circle*{3}}
\put(-4,-8){\circle*{3}}
\put(5,-8){\circle*{3}}
\put(10,2){\circle*{3}}
\put(5,8){\circle*{3}}
\put(0,0){\line(0,1){15}}
\put(0,0){\line(5,-8){5}}
\put(0,0){\line(-5,1){10}}
\put(-10,2){\line(6,-10){6}}
\put(5,-8){\line(1,2){5}}
\put(5,8){\line(0,-1){16}}
\qbezier(-6,8)(2,35)(10,2)
\qbezier(-4,-8)(-25,4)(0,15)
\end{picture}
&\math{\rightarrow}&
\begin{picture}(16,30)(-6,-8)
\put(0,0){\circle*{3}}
\put(0,15){\circle*{3}}
\put(-6,8){\circle*{3}}
\put(-10,2){\circle*{3}}
\put(-4,-8){\circle*{3}}
\put(5,-8){\circle*{3}}
\put(10,2){\circle*{3}}
\put(5,8){\circle*{3}}
\put(0,0){\line(0,1){15}}
\put(0,0){\line(5,-8){5}}
\put(-10,2){\line(6,-10){6}}
\put(5,-8){\line(1,2){5}}
\put(-10,2){\line(3,-2){15}}
\put(5,8){\line(-5,-8){5}}
\qbezier(-6,8)(2,35)(10,2)
\qbezier(-4,-8)(-25,4)(0,15)
\end{picture}
&\math{\rightarrow}&
\begin{picture}(16,30)(-6,-8)
\put(0,0){\circle*{3}}
\put(0,15){\circle*{3}}
\put(-6,8){\circle*{3}}
\put(-10,2){\circle*{3}}
\put(-4,-8){\circle*{3}}
\put(5,-8){\circle*{3}}
\put(10,2){\circle*{3}}
\put(5,8){\circle*{3}}
\put(0,0){\line(5,-8){5}}
\put(-10,2){\line(6,-10){6}}
\put(5,-8){\line(1,2){5}}
\put(0,15){\line(-6,-7){6}}
\put(-10,2){\line(3,-2){15}}
\put(5,8){\line(-5,-8){5}}
\put(0,0){\line(5,1){10}}
\qbezier(-4,-8)(-25,4)(0,15)
\end{picture}
\\[2pt]
\math{N_0}
&&
\math{N_1}
&&
\math{N_2}
&&
\math{N_3}
&&
\math{N_4}
\end{tabular}}%

\noindent
Random edge swaps slowly dismantle the ``structure'',
yielding, in the limit, a random graph with the same degrees as~\math{N_0}.

Let \math{W} be a random process that visits vertices.
For concreteness,
from now on \math{W} is a random walk which 
traverses a random incident edge at each step. After \math{W}
visits~\math{\kappa} different vertices, construct
the subgraph induced by those \math{\kappa} vertices:
\mldc{
W:(N_\delta,\kappa)\mapsto G(\kappa,\delta),
}
where \math{G(\kappa,\delta)} is a random graph that depends on
the network \math{N_\delta}, the start vertex and edges
traversed. The process
\math{W} induces a distribution
\math{p_{\kappa,\delta}} over graphs with~\math{\kappa} vertices.
If the distributions
\math{p_{\kappa,0}} and \math{p_{\kappa,\delta}} are distinguishable, 
existing structure in \math{N=N_0} at scale~\math{\kappa} was lost
during the \math{\delta} steps of randomization that produced~\math{N_\delta}.
The Bayes \emph{optimal} classifier for the distributions
\math{p_{\kappa,0}} and \math{p_{\kappa,\delta}} has classification
accuracy
\mldc{
\Delta(\kappa,\delta)={\textstyle\frac12\sum\limits_{G}}
\max\{p_{\kappa,0}(G), p_{\kappa,\delta}(G)\}.
\label{eq:bayes-optimal}
}
We focus on \math{\delta\rightarrow\infty},
in which case, \math{N_\delta}
is a random graph with the same
vertex-degrees as \math{N}.
If \math{\Delta(\kappa,\infty)\gg \frac12},
 one can distinguish \math{\kappa}-sized subgraphs of
\math{N} form those in \math{N_\infty} with high accuracy,
which means there is significant
structure at the scale \math{\kappa} in \math{N}. Hence, we define the
intrinsic
scale \math{\kappa^*(\tau)}:
\begin{definition}[Intrinsic Scale]\label{def:scale}
For \math{\tau>\frac12}, let 
\math{\kappa^*(\tau,\delta)} be the minimum scale \math{\kappa}
at which one can distinguish \math{\kappa}-sized subgraphs of
\math{N} from those in \math{N_\delta} with accuracy at least \math{\tau},
\mldc{\kappa^*(\tau,\delta)=\min\{\kappa\mid
\Delta(\kappa,\delta)\ge\tau\}.
\label{eq:scale-def}}
The intrinsic scale is
\math{\kappa^*(\tau)=\lim_{\delta\rightarrow\infty}\kappa^*(\tau,\delta)}.
\end{definition}
Implicit in the definition of intrinsic scale
is the process \math{W} which produces
\math{\kappa}-sized subgraphs. The details of \math{W} can affect
specific values of \math{\kappa^*}, and it is natural to focus on
subgraphs which are ``locally'' constructed as with a random walk.

\begin{example}{ (Intrinsic scale of trees)}\label{example:tree}
We show a 5-node labeled tree~\math{N_0} in the figure below (leftmost).
Edge swapping
will
randomly produce one of the 8 graphs shown (note, we allow
parallel edges).
\vskip4pt

\centerline{\tabcolsep1pt\setlength{\unitlength}{1.15pt}\thicklines%
\hspace*{-1pt}\begin{tabular}{cccccccc}
\fboxsep1pt\fbox{%
\begin{picture}(23,18)(-7,-2)
\put(0,0){\circle*{3}}
\put(3.5,7){\circle*{3}}
\put(7,14){\circle*{3}}
\put(10.5,7){\circle*{3}}
\put(14,0){\circle*{3}}
\put(0,0){\line(1,2){3.5}}
\put(3.5,7){\line(1,2){3.5}}
\put(7,14){\line(1,-2){3.5}}
\put(10.5,7){\line(1,-2){3.5}}
\put(-8,10){\footnotesize\math{N_0}}
\end{picture}%
}
&
\fboxsep1pt\fbox{%
\begin{picture}(18,18)(-2,-2)
\put(0,0){\circle*{3}}
\put(3.5,7){\circle*{3}}
\put(7,14){\circle*{3}}
\put(10.5,7){\circle*{3}}
\put(14,0){\circle*{3}}
\put(0,0){\line(1,2){3.5}}
\put(3.5,7){\line(3,-2){11.5}}
\qbezier(7,14)(14,13.125)(10.5,7)
\qbezier(7,14)(3.5,7.875)(10.5,7)
\end{picture}%
}
&
\fboxsep1pt\fbox{%
\begin{picture}(18,18)(-2,-2)
\put(0,0){\circle*{3}}
\put(3.5,7){\circle*{3}}
\put(7,14){\circle*{3}}
\put(10.5,7){\circle*{3}}
\put(14,0){\circle*{3}}
\put(0,0){\line(3,2){11.5}}
\put(10.5,7){\line(1,-2){3.5}}
\qbezier(3.5,7)(0,13.125)(7,14)
\qbezier(3.5,7)(10.5,7.875)(7,14)
\end{picture}%
}
&
\fboxsep1pt\fbox{%
\begin{picture}(18,18)(-2,-2)
\put(0,0){\circle*{3}}
\put(3.5,7){\circle*{3}}
\put(7,14){\circle*{3}}
\put(10.5,7){\circle*{3}}
\put(14,0){\circle*{3}}
\put(0,0){\line(1,2){3.5}}
\put(3.5,7){\line(1,0){7}}
\qbezier(7,14)(14,14)(14,0)
\put(10.5,7){\line(-1,2){3.5}}
\end{picture}%
}
&
\fboxsep1pt\fbox{%
\begin{picture}(18,18)(-2,-2)
\put(0,0){\circle*{3}}
\put(3.5,7){\circle*{3}}
\put(7,14){\circle*{3}}
\put(10.5,7){\circle*{3}}
\put(14,0){\circle*{3}}
\put(14,0){\line(-1,2){3.5}}
\put(3.5,7){\line(1,0){7}}
\qbezier(7,14)(0,14)(0,0)
\put(3.5,7){\line(1,2){3.5}}
\end{picture}%
}
&
\fboxsep1pt\fbox{%
\begin{picture}(18,18)(-2,-2)
\put(0,0){\circle*{3}}
\put(3.5,7){\circle*{3}}
\put(7,14){\circle*{3}}
\put(10.5,7){\circle*{3}}
\put(14,0){\circle*{3}}
\put(0,0){\line(3,2){10.5}}
\put(3.5,7){\line(1,2){3.5}}
\put(7,14){\line(1,-2){3.5}}
\put(3.5,7){\line(3,-2){10.5}}
\end{picture}%
}
&
\fboxsep1pt\fbox{%
\begin{picture}(18,18)(-2,-2)
\put(0,0){\circle*{3}}
\put(3.5,7){\circle*{3}}
\put(7,14){\circle*{3}}
\put(10.5,7){\circle*{3}}
\put(14,0){\circle*{3}}
\put(0,0){\line(1,0){14}}
\put(3.5,7){\line(1,0){7}}
\put(7,14){\line(1,-2){3.5}}
\put(3.5,7){\line(1,2){3.5}}
\end{picture}%
}
&
\fboxsep1pt\fbox{%
\begin{picture}(18,18)(-2,-2)
\put(0,0){\circle*{3}}
\put(3.5,7){\circle*{3}}
\put(7,14){\circle*{3}}
\put(10.5,7){\circle*{3}}
\put(14,0){\circle*{3}}
\qbezier(3.5,7)(7,14)(10.5,7)
\qbezier(3.5,7)(7,0)(10.5,7)
\qbezier(7,14)(0,14)(0,0)
\qbezier(7,14)(14,14)(14,0)
\end{picture}%
}
\end{tabular}%
}%
\vskip4pt

\noindent
Connectivity alone distinguishes between the original
tree and the perturbed graph with 50\% accuracy. When the tree
size increases, the accuracy improves.
\\[-8pt]

\centerline{
	\includegraphics[width=0.3\textwidth]{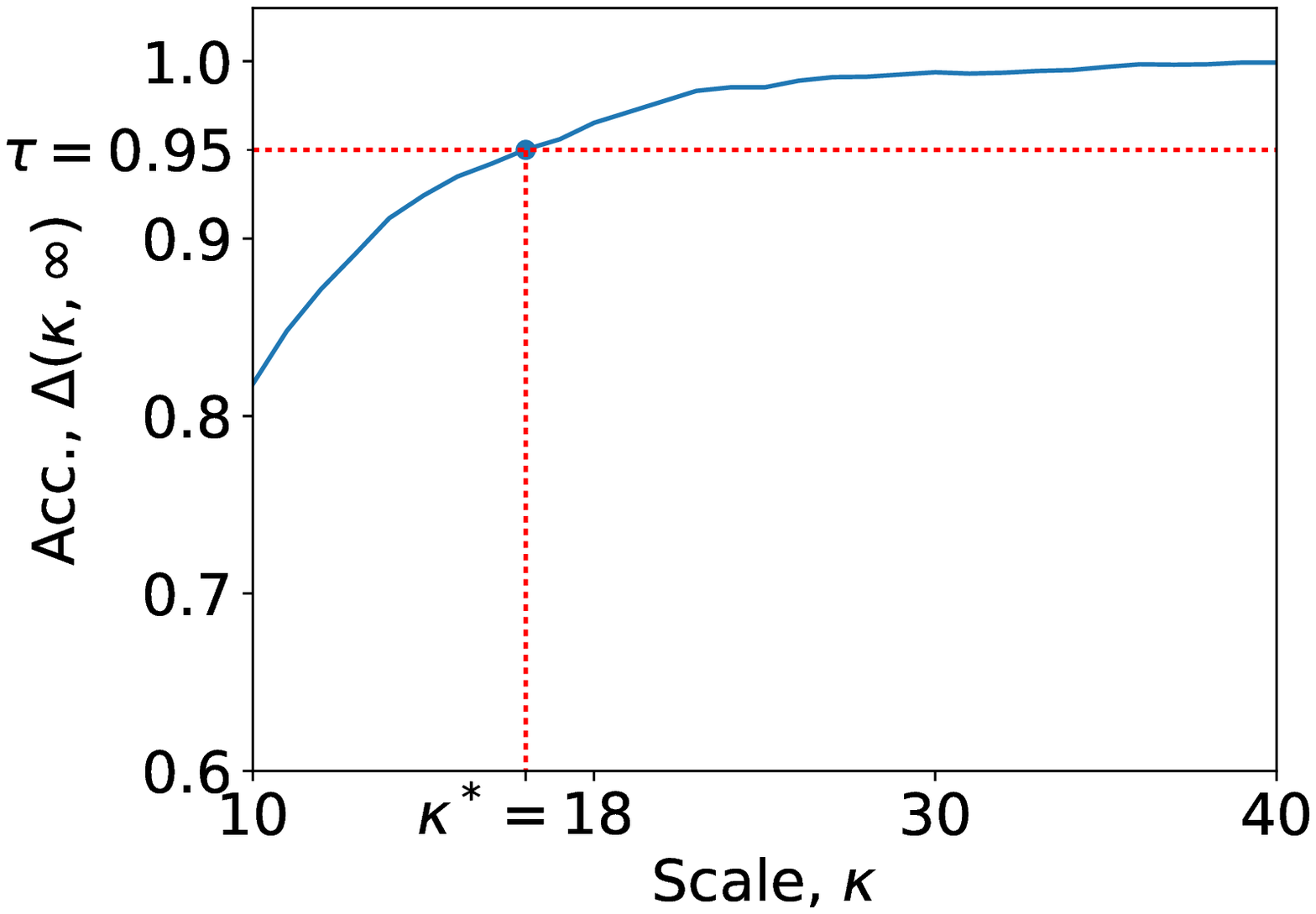}
}
\noindent
With 95\% accuracy, a random tree of size 18 can be distinguished from
a random graph with the same degrees.
The intrinsic scale at 95\% accuracy is \math{\kappa^*\le 18}
(upper bound because we are not using the Bayes optimal
classifier, just one based on connectivity).
\end{example}%
The example hits an important point.
As the subgraph-size~\math{\kappa}  increases, 
computing the accuracy in (\ref{eq:bayes-optimal})
is exponential.
To make the computation feasible, we summarized a subgraph using
a statistic,
connectivity,
and obtained the classification accuracy using just that
feature. This only  gives a lower bound on the optimal accuracy.
The same statistic may not work for every type of network.
For example, with a large clique, random edge swaps would still maintain
connectivity, and some other discriminative
statistic would be needed to avoid the exponential complexity
in (\ref{eq:bayes-optimal}).

Our notion of structure at scale \math{\kappa} corresponds to a
game where I show you a random \math{\kappa}-sized subgraph and ask if you
are surprised. You will be surprised if you see some
``unexpected'' structure. A 20-node
clique might surprise \emph{you} because
you have an internal null distribution
for random graphs, from which 
a 20-node clique is  unlikely -- 
has ``unexpected'' structure.
We define this null distribution concretely as \math{p_{\kappa,\infty}},
which is natural as it is non informative over graphs with the
same degrees. Our methodology, however,
works with any other way to  construct the null distribution
while preserving desired properties of the graph
(see for example~\cite{Mukherjee14313}).

We summarize
our main findings in Table~\ref{tab:kappa} (the
\math{\delta=\infty} column), which
gives upper bounds on the intrinsic
scale of some real networks.
Even
at  \math{95\%} accuracy, the intrinsic scale of
real networks is no more than 20, for small and 
large networks alike.
Traditionally structured networks, like roads, have smaller
intrinsic scale (no surprise), while
loosely structured networks like Wikipedia have larger
intrinsic scale. Interestingly, the biological protein networks have
comparatively large intrinsic scale, which indicates they have less structure
than one might expect, perhaps due to the need for degeneracy,
redundancy and robustness, \cite{Tononi3257}.
At 70\% accuracy, almost all
networks have structure at very small scales that is fragile and
easily disrupted with just 30\% of
edge-swaps.

\begin{table}
\centering
{%
{\tabcolsep5pt
\begin{tabular}{c||c|c|c||c|c|c||c|c|c}
\multirow{2}{*}{$\kappa^{*}(\tau)$}& \multicolumn{3}{c||}{$\tau=0.7$}  & \multicolumn{3}{c||}{$\tau=0.9$} & \multicolumn{3}{c}{$\tau=0.95$}\tabularnewline
 & 30 & 50 & $\infty$ & 30 & 50 & $\infty$ & 30 & 50 & \textbf{$\infty$}\tabularnewline
\hline 
Road
&8 &7 & \textbf{4}
& 24 & 16 & \textbf{4}
& 34 & 22 & \textbf{7}
\tabularnewline
Facebook
& 4 & 4& \textbf{4}
& 13 & 10 & \textbf{8}
& 18 & 15 & \textbf{10}
\tabularnewline
Human
& 28& 21& \textbf{4}
& $*$ & $*$ & \textbf{6}
& $*$ & $*$ & \textbf{10}
\tabularnewline
Amazon
& 5& 4 & \textbf{4}
& 24 & 14 & \textbf{9}
& 50 & 24 & \textbf{12}
\tabularnewline
Al-Qaeda
&4 &4 &\textbf{4}
& 14 & 10 & \textbf{9}
& 21 & 15 & \textbf{12}
\tabularnewline
Cite
& 8& 6& \textbf{4}
& 32 & 20 & \textbf{10}
& 64 & 33 & \textbf{12}
\tabularnewline
DBLP
& 4& 4& \textbf{4}
& 27 & 15 & \textbf{9}
& 52 & 26 & \textbf{13}
\tabularnewline
Web
& 7& 5&\textbf{4}
& 36 & 21 & \textbf{7}
& 64 & 37 & \textbf{14}
\tabularnewline
Gowalla
& 16& 11 & \textbf{7}
& $*$ & $*$ & \textbf{14}
& $*$ & $*$ & \textbf{17}
\tabularnewline
Mouse
& 28& 22& \textbf{4}
& $*$ & 62 & \textbf{15}
& $*$ & $*$ & \textbf{20}
\tabularnewline
Yeast
&28 & 18& \textbf{5}
& $*$ & $*$ & \textbf{15}
& $*$ & $*$ & \textbf{20}
\tabularnewline
Wiki
& 48 & 32& \textbf{9}
& $*$ & $*$ & \textbf{16}
& $*$ & $*$ & \textbf{20}
\tabularnewline
\end{tabular}}}
\caption{Intrinsic scale of networks for $\delta \in \{30\%, 50\%, \infty\}$.
We use 
$*$ to mean \math{\kappa^*(\tau,\delta)>64} (the maximum size in our
experiments), which means that while there might be structure,
it was robust to the perturbation
(couldn't be systematically discriminated).
\label{tab:kappa}}
\end{table}

\section{Data and Methods}
\label{datamethods}

We tested a variety of networks (see Table \ref{tab:data}), and all graph algorithms were implemented in Python using NetworkX \cite{hagberg2008}.

\begin{table*}
\centering
\begin{tabular}{c|c|c|c}
Network& Type & \# Nodes, \math{n} & \# Edges, \math{m}\tabularnewline
\hline 
Road \cite{roadLeskovec2009} &Infrastructure& 1,088,092 & 1,541,898\tabularnewline
Web \cite{roadLeskovec2009} &Information& 875,713 & 5,105,039\tabularnewline
Amazon \cite{amazonLeskovec2007}&e-Commerce & 334,863 & 925,872\tabularnewline
DBLP \cite{dblpYang2012}&Citation & 317,080 & 1,049,866\tabularnewline
Gowalla \cite{cho2011friendship}&Social  & 196,591 & 950,327\tabularnewline
Citation \cite{citationLeskovec2005,citationGehrke2003}&Citation & 34,546 & 421,578\tabularnewline
Human \cite{reimand2008graphweb}&PPI  & 8,077 & 26,085\tabularnewline
Yeast \cite{reimand2008graphweb}&PPI  & 5,718 & 48,253\tabularnewline
Wiki \cite{wikiWest2012,wiki2West2009}&Information & 4,604 & 119,882\tabularnewline
Facebook \cite{facebookMcAuley2012} &Social& 4,039 & 88,234\tabularnewline
Mouse \cite{reimand2008graphweb}&PPI & 2,929 & 4,188\tabularnewline
Al-Qaeda \cite{jjatt}&Social & 271 & 756\tabularnewline
\end{tabular}
\caption{Datasets used in this study}
\label{tab:data}
\end{table*}
The main challenge is to efficiently estimate
the Bayes optimal accuracy
\math{\Delta(\kappa,\delta)}
in \r{eq:bayes-optimal}, without computing the full distributions
\math{p_{\kappa,0}} and \math{p_{\kappa,\delta}}.
Given \math{\Delta(\kappa,\delta)}, we compute the intrinsic
scale using \r{eq:scale-def}.
Our approach to computing \math{\Delta(\kappa,\delta)}
is to sample subgraphs and formulate the task as
a standard machine learning problem. The workflow is as follows.
\\[-4pt]

\begin{algorithm}[tb]
\caption{Estimate \math{\Delta(\kappa,\delta)}}
\label{page:alg1}
\begin{algorithmic}
	\STATE{1.} Given \math{N_0}, construct \math{N_\delta} using \math{\delta} edge-swaps.
\STATE{2.} Sample \math{\kappa}-sized subgraphs from \math{N_0} and \math{N_\delta} to get a training set\math{\{G_{\kappa,0}\}^{\text{train}}} and \math{\{G_{\kappa,\delta}\}^{\text{train}}}.
\STATE{3.} Use the training set to learn a classifier\mldc{g_{\kappa,\delta}: G_\kappa\mapsto\pm1.} (\math{+1} for \math{N_0}, \math{-1} for \math{N_\delta}).
\STATE{4.} Test the learned classifier \math{g_{\kappa,\delta}} on independent  test subgraphs \math{\{G_{\kappa,0}\}^{\text{test}}} and \math{\{G_{\kappa,\delta}\}^{\text{test}}}.
\STATE{5.} Return \math{\hat\Delta(\kappa,\delta)}, the test accuracy of \math{g_{\kappa,\delta}}.
\end{algorithmic}
\end{algorithm}
\vspace*{4pt}

\noindent
In step 1,
edge-swaps preserve vertex degrees. For \math{\delta=\infty},
\math{N_\delta} is a random graph with the same degrees
as \math{N_0}. In steps 2 and~4, the training and test graphs
are sampled using the random walker~\math{W}. A larger training set
gives a better learned classifier \math{g_{\kappa,\delta}}; a larger test set
gives a better accuracy-estimate for \math{g_{\kappa,\delta}}.
We used 10,000 samples from each graph, half
for training and the rest for test. The Bayes
optimal accuracy for the classification problem is
\math{\Delta(\kappa,\delta)\ge\hat\Delta(\kappa,\delta)}. The best
estimate of \math{\Delta(\kappa,\delta)} comes from best
learned classifier \math{g_{\kappa,\delta}},
hence the learning algorithm is important.

\begin{figure}[ht!]
  \centering
{\tabcolsep2pt
\begin{tabular}{cccccc}
&{$N_0$}&
{$N_{10\%}$}&
{$N_{30\%}$}&
{$N_{50\%}$}&
{$N_\infty$}\\
  \rotatebox{90}{\footnotesize Facebook, \math{64}}&\fboxsep0pt\fbox{\includegraphics[scale=0.19]{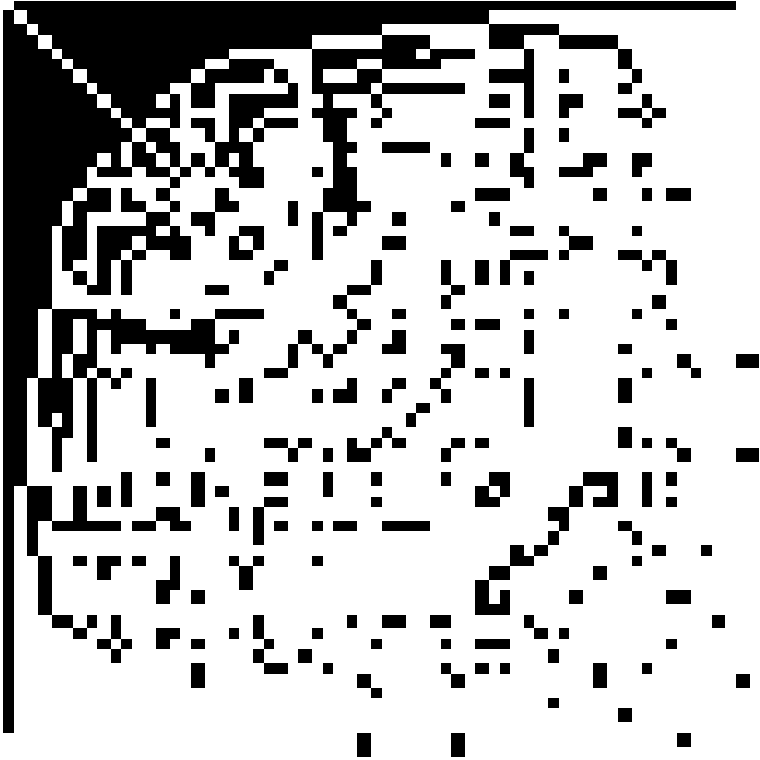}}&
  \fboxsep0pt\fbox{\includegraphics[scale=0.19]{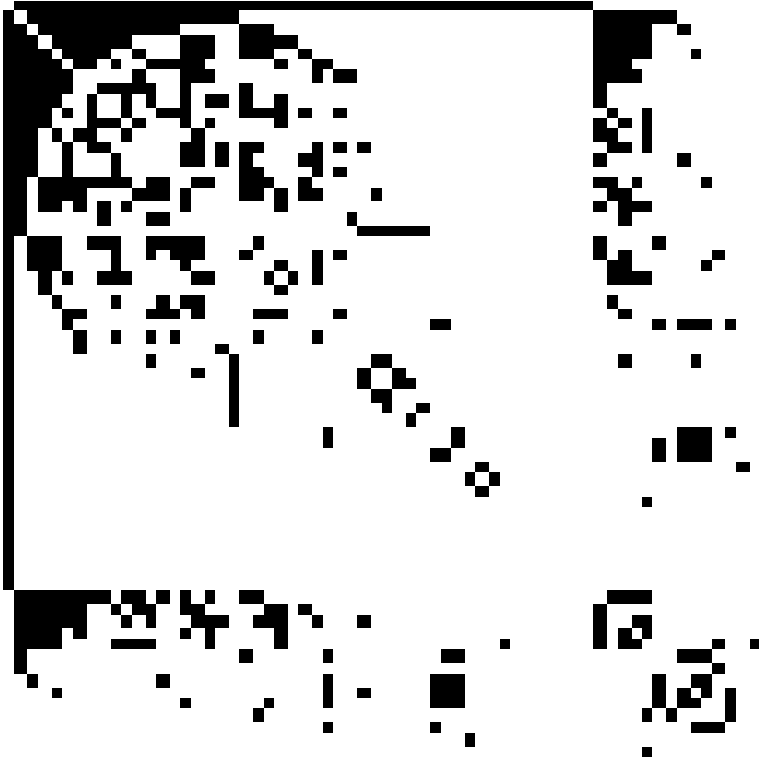}}&
  \fboxsep0pt\fbox{\includegraphics[scale=0.19]{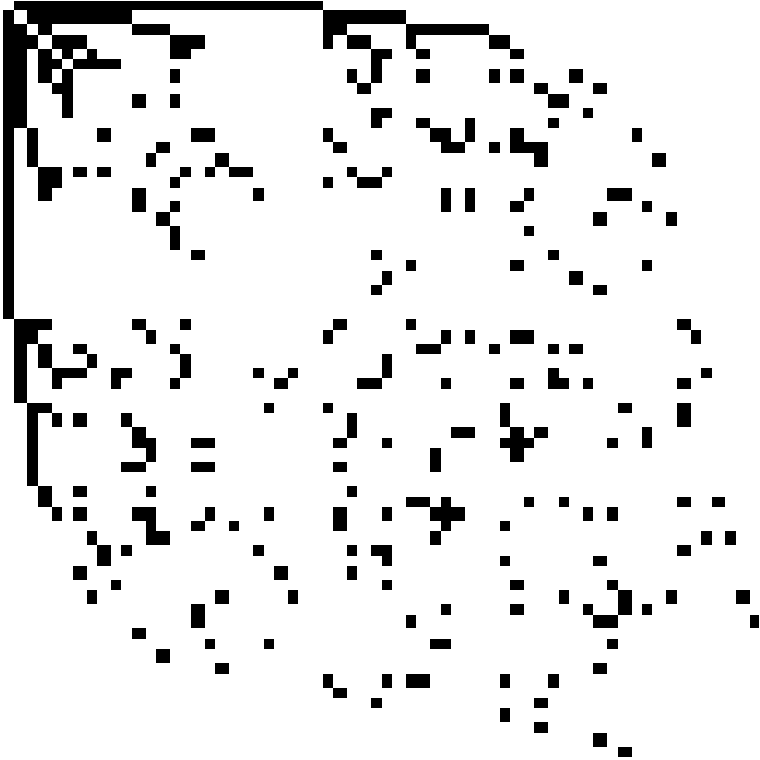}}&
  \fboxsep0pt\fbox{\includegraphics[scale=0.19]{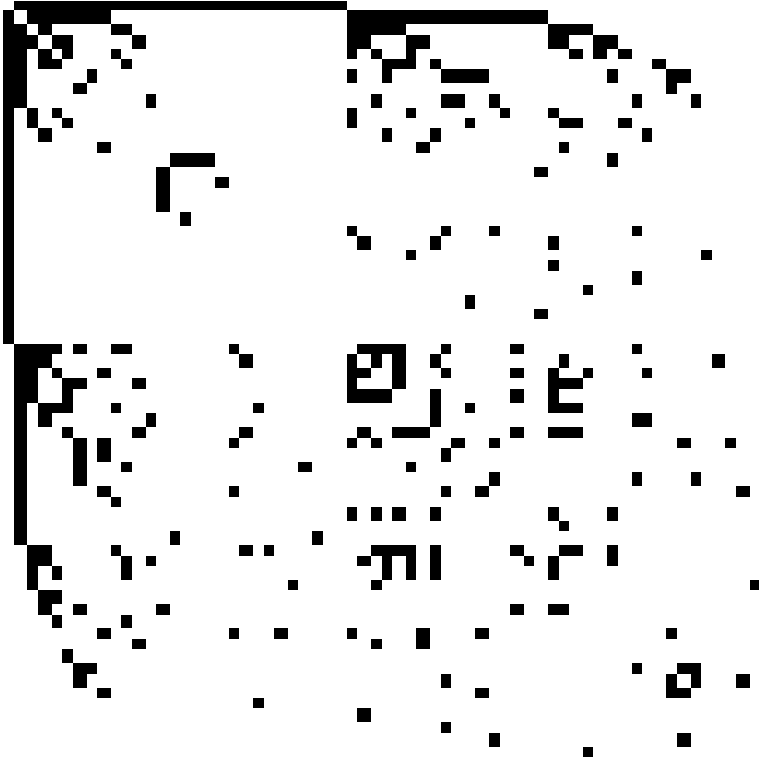}}&
  \fboxsep0pt\fbox{\includegraphics[scale=0.19]{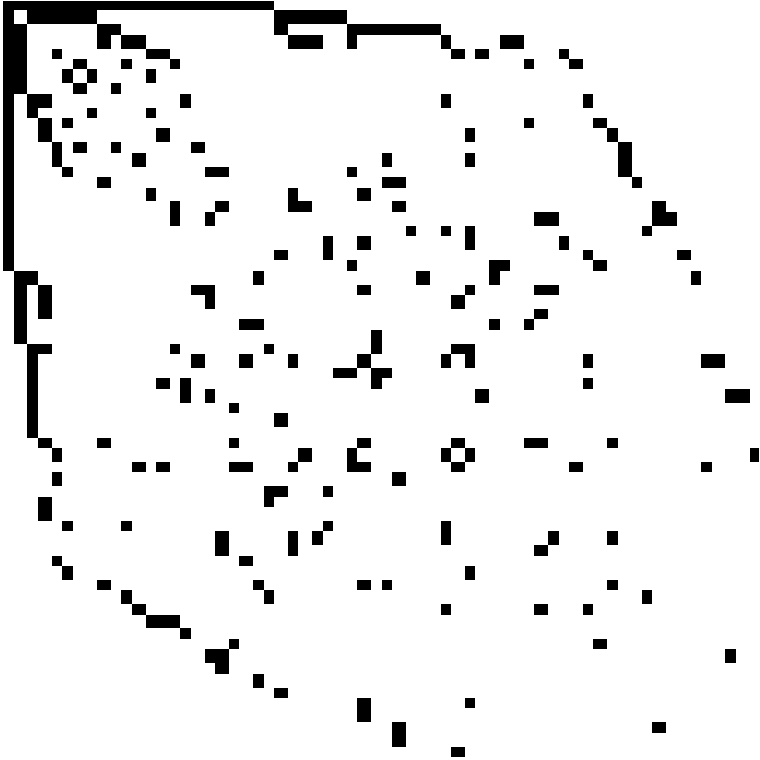}}\\
  \rotatebox{90}{\footnotesize Wikipedia, \math{64}}&\fboxsep0pt\fbox{\includegraphics[scale=0.19]{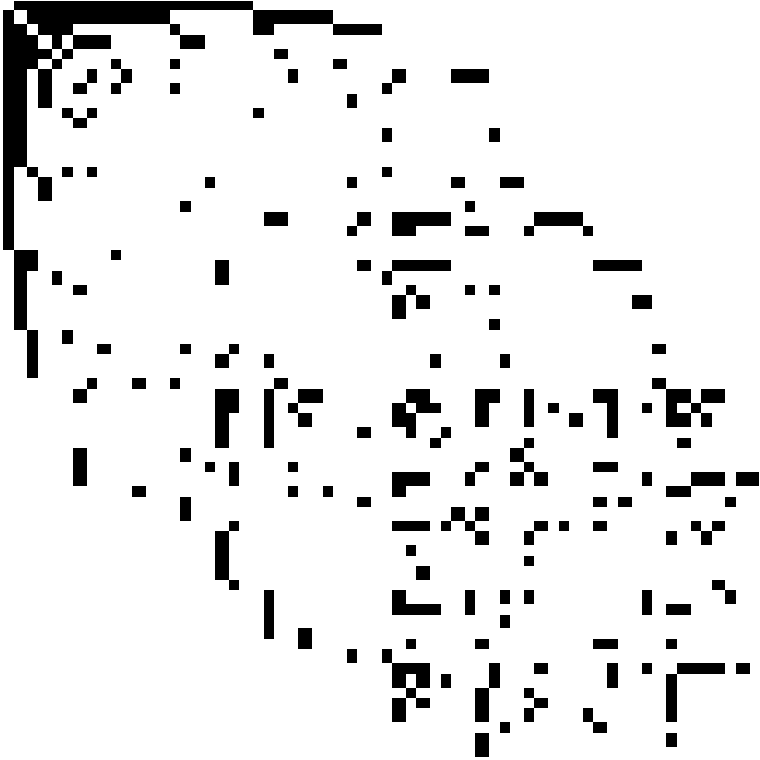}}&
  \fboxsep0pt\fbox{\includegraphics[scale=0.19]{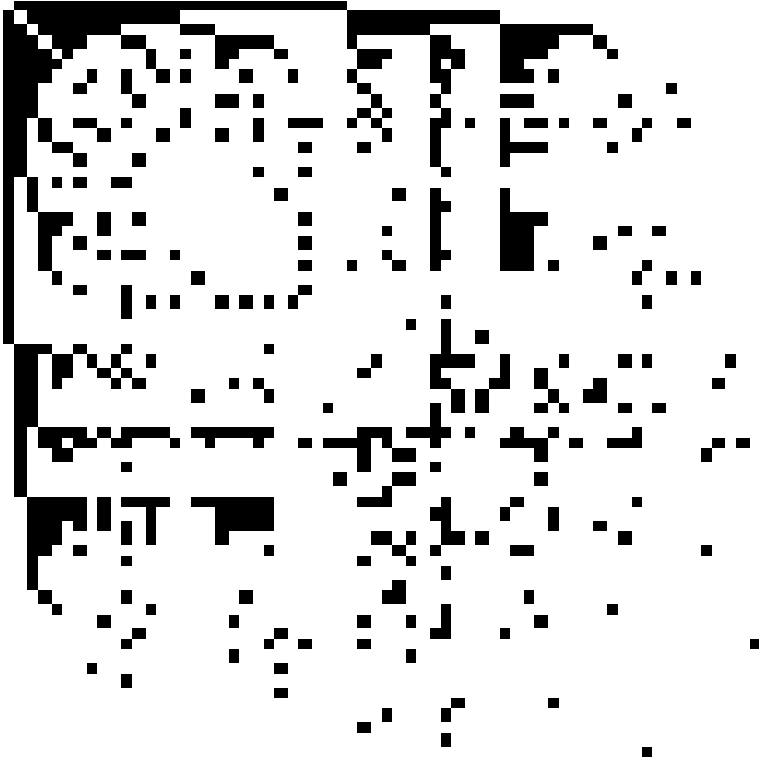}}&
  \fboxsep0pt\fbox{\includegraphics[scale=0.19]{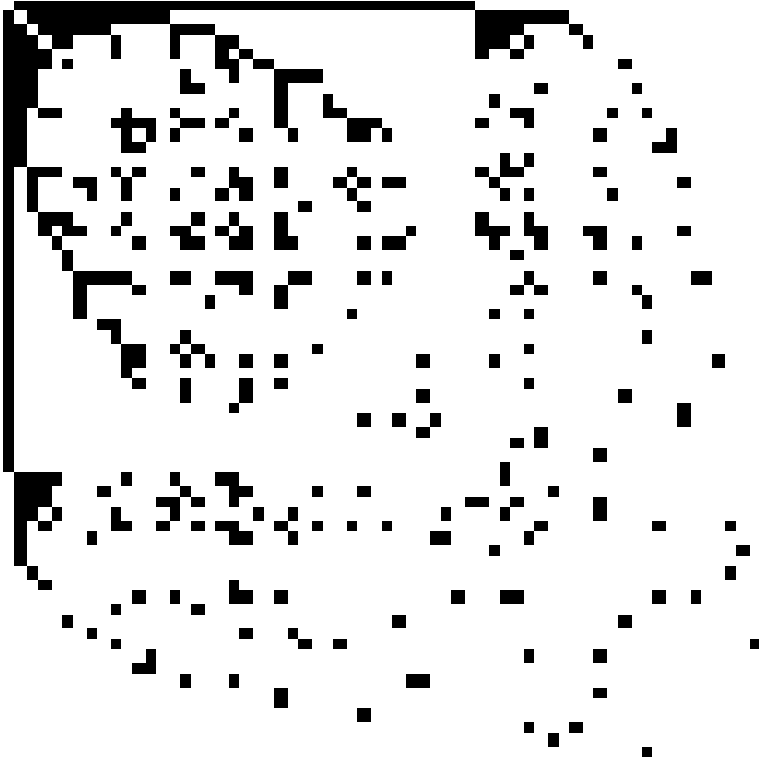}}&
  \fboxsep0pt\fbox{\includegraphics[scale=0.19]{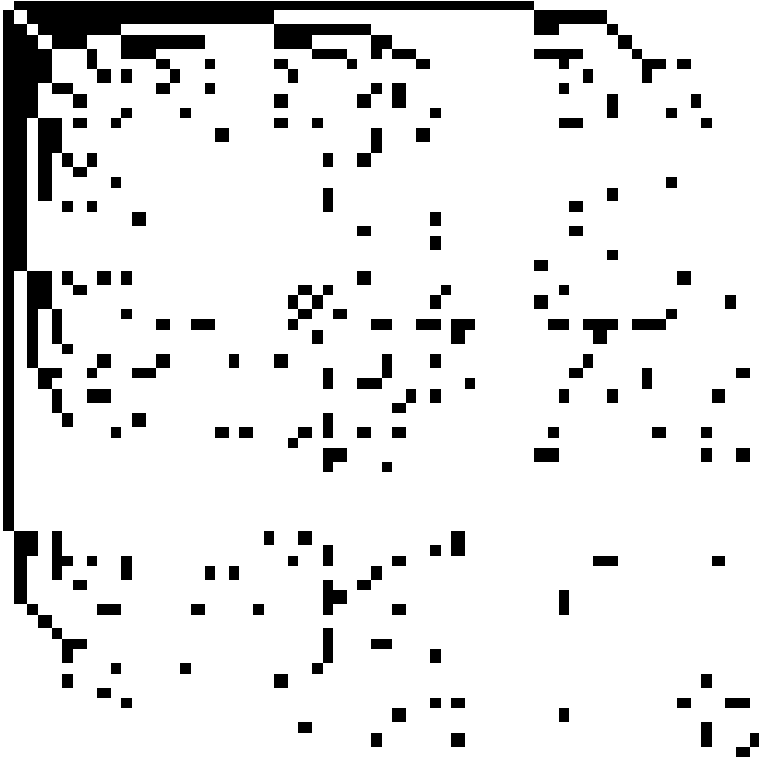}}&
  \fboxsep0pt\fbox{\includegraphics[scale=0.19]{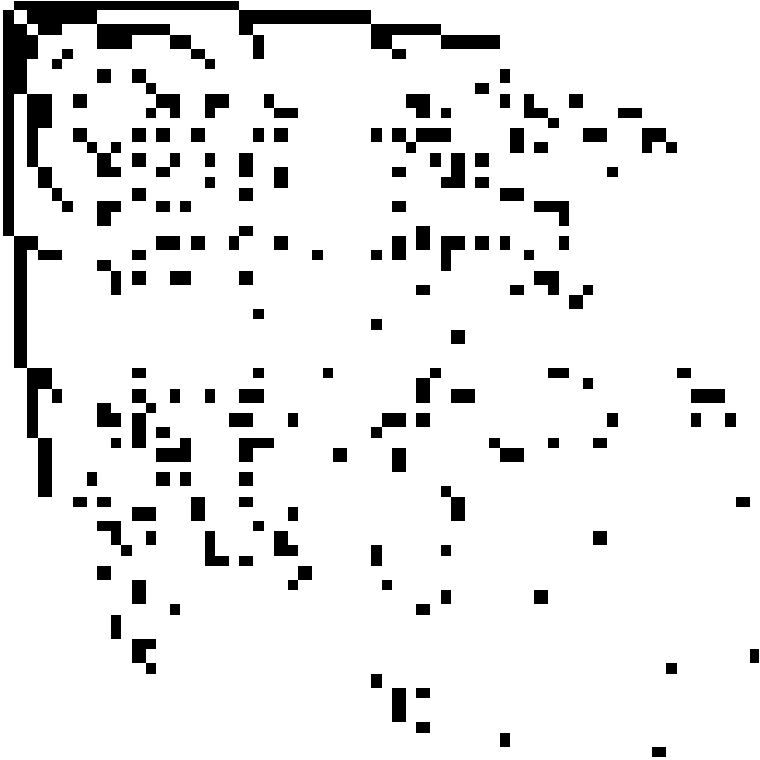}}
  \\[3pt]
  \hline
  \\[-5pt]
  \rotatebox{90}{\footnotesize Facebook, \math{16}}&\fboxsep-2pt\fbox{\includegraphics[scale=0.18]{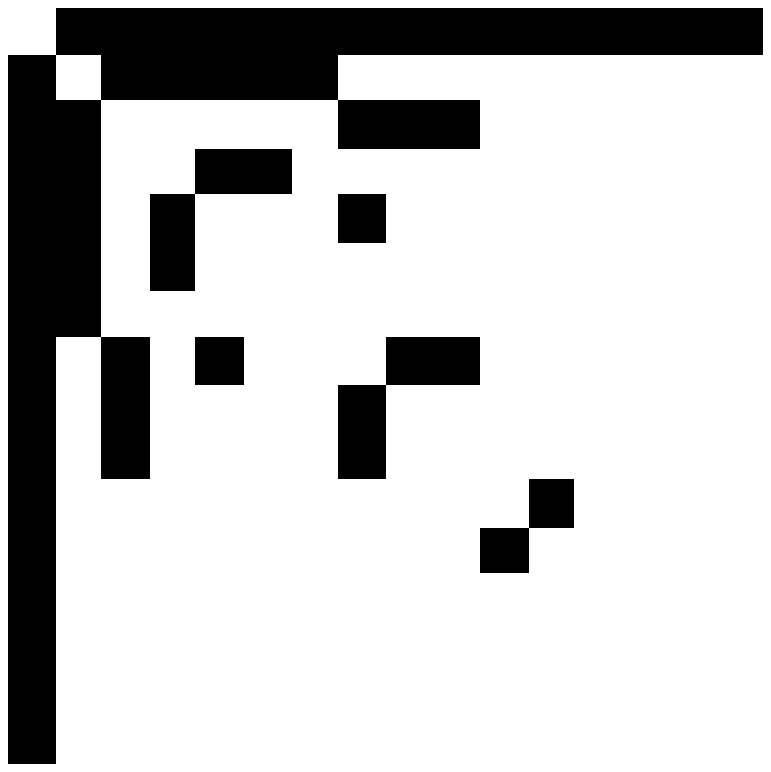}}&
  \fboxsep-2pt\fbox{\includegraphics[scale=0.18]{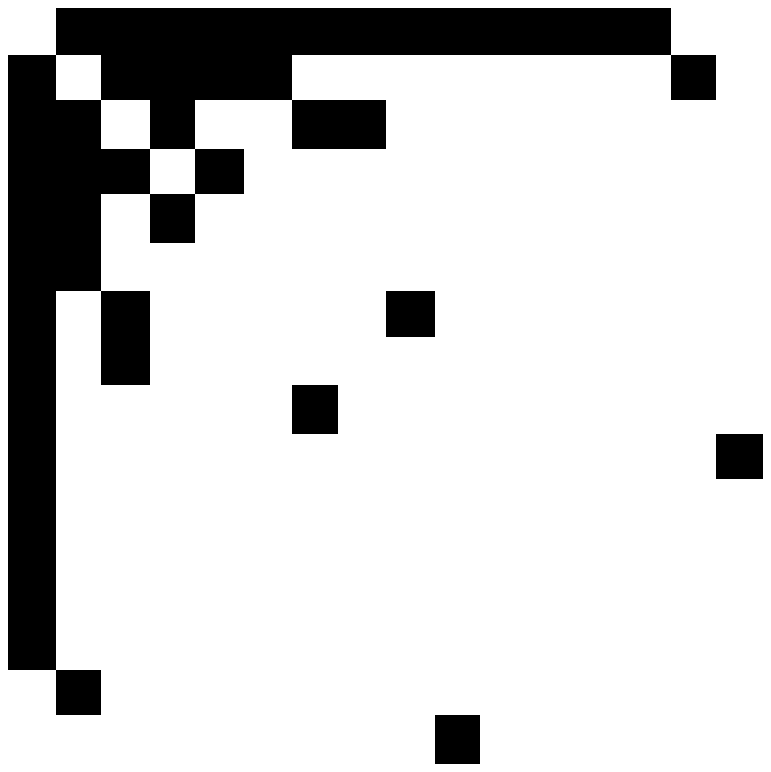}}&
  \fboxsep-2pt\fbox{\includegraphics[scale=0.18]{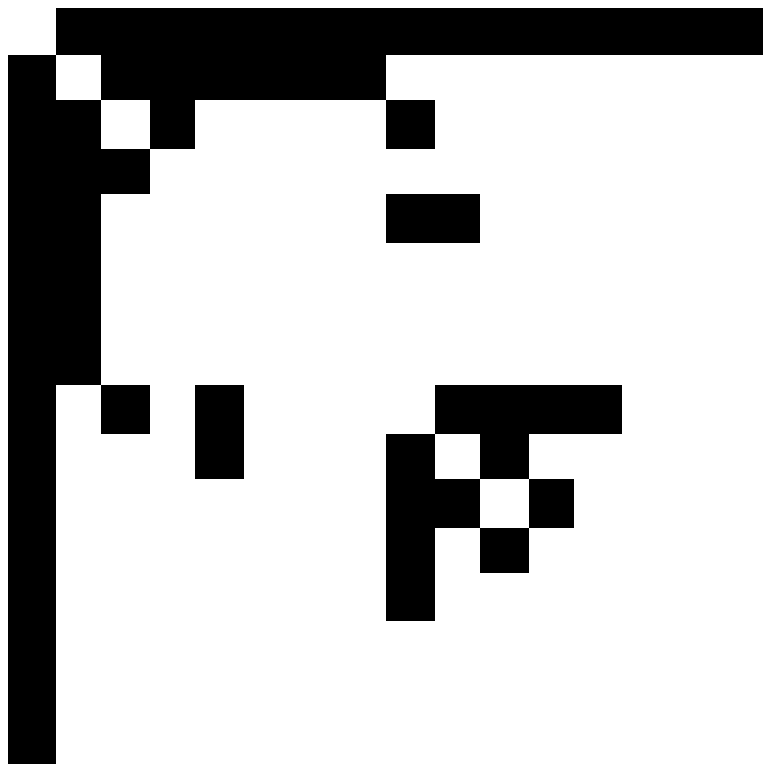}}&
  \fboxsep-2pt\fbox{\includegraphics[scale=0.18]{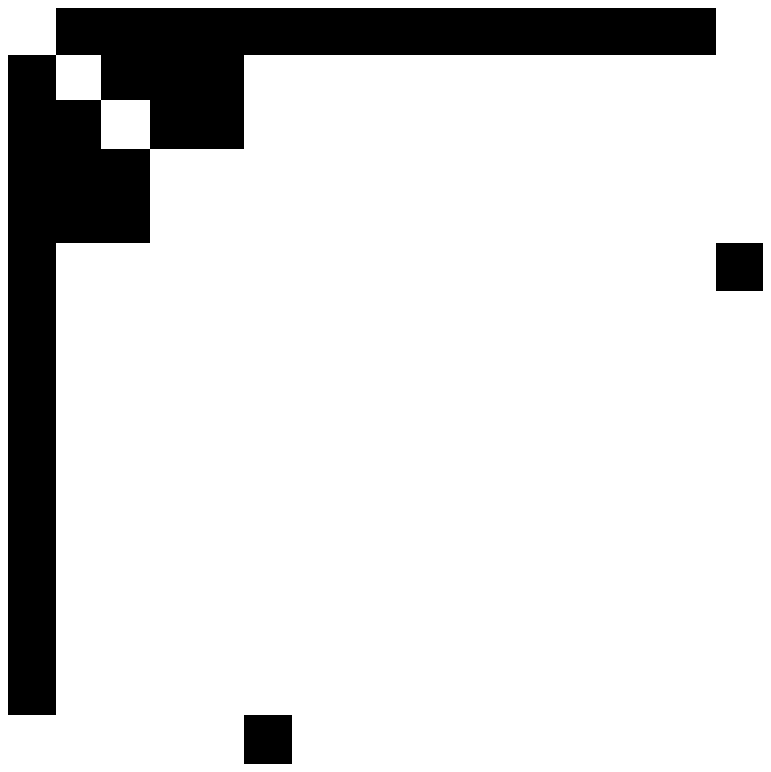}}&
  \fboxsep-2pt\fbox{\includegraphics[scale=0.18]{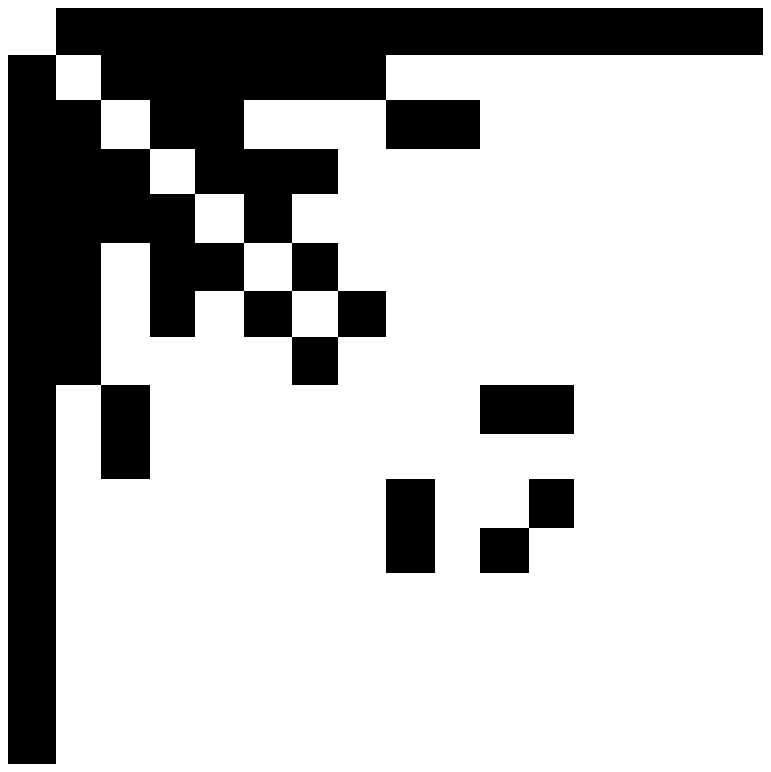}}\\
  \rotatebox{90}{\footnotesize Wikipedia, \math{16}}&\fboxsep-2pt\fbox{\includegraphics[scale=0.18]{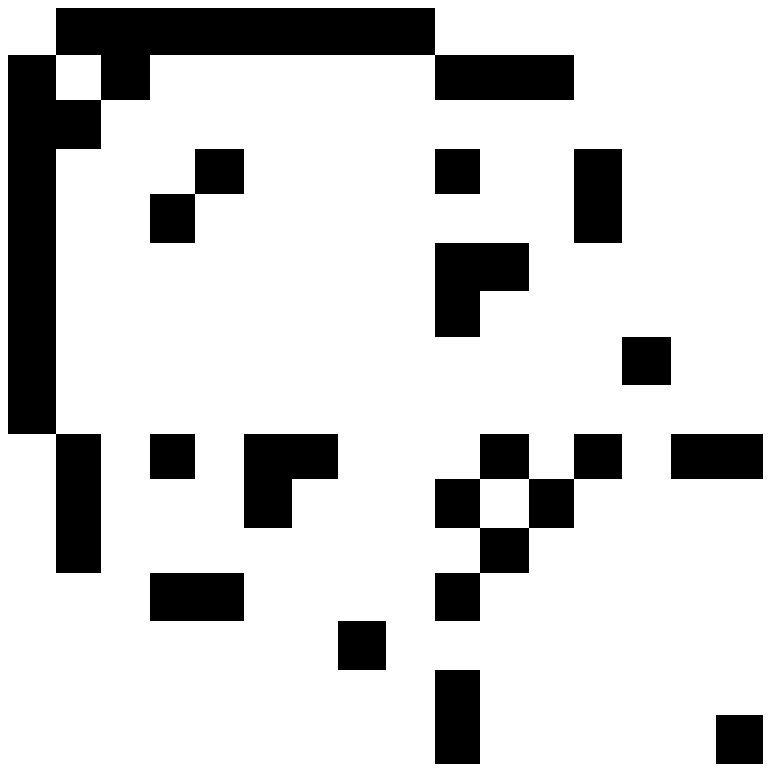}}&
  \fboxsep-2pt\fbox{\includegraphics[scale=0.18]{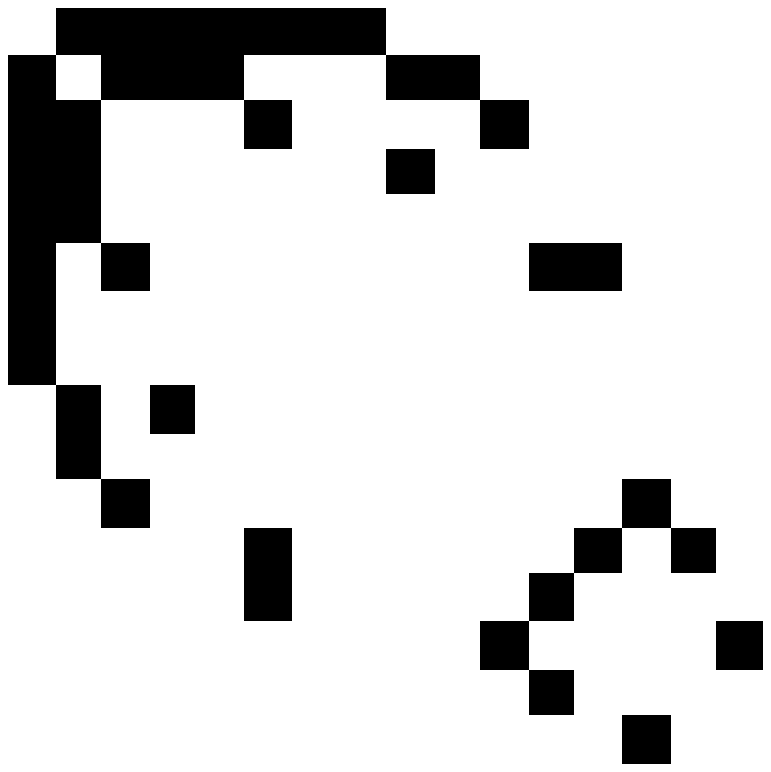}}&
  \fboxsep-2pt\fbox{\includegraphics[scale=0.18]{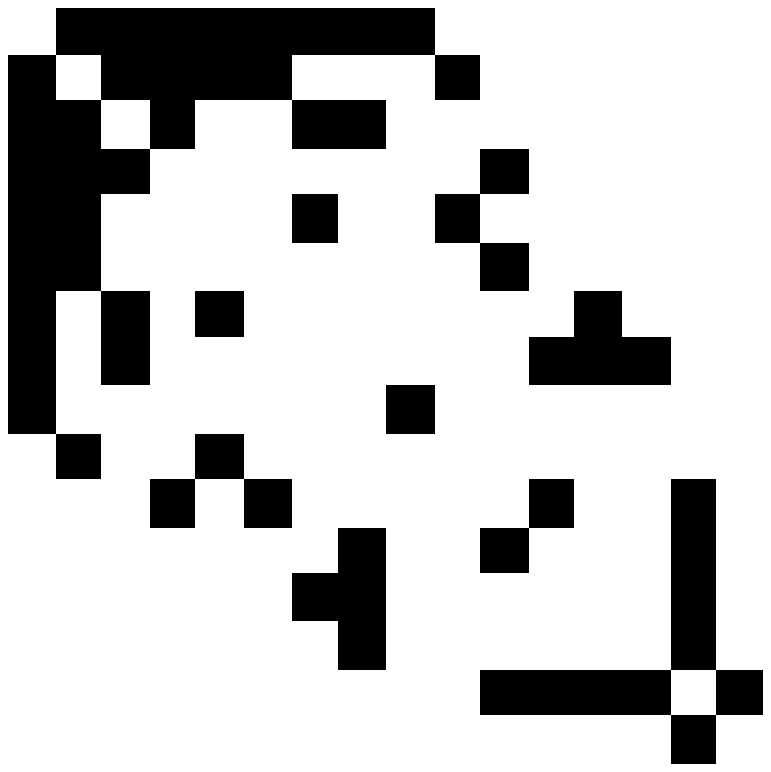}}&
  \fboxsep-2pt\fbox{\includegraphics[scale=0.18]{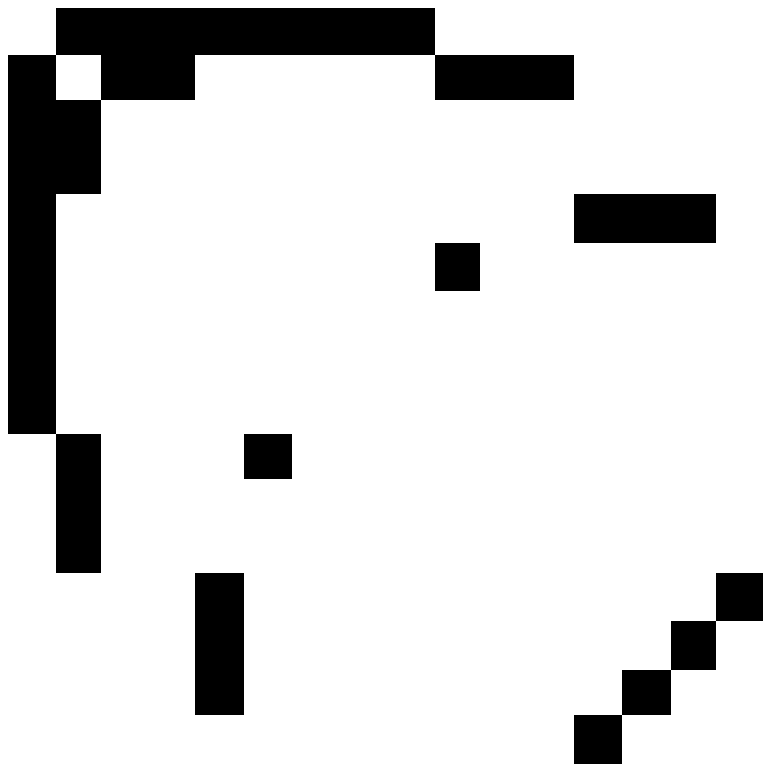}}&
  \fboxsep-2pt\fbox{\includegraphics[scale=0.18]{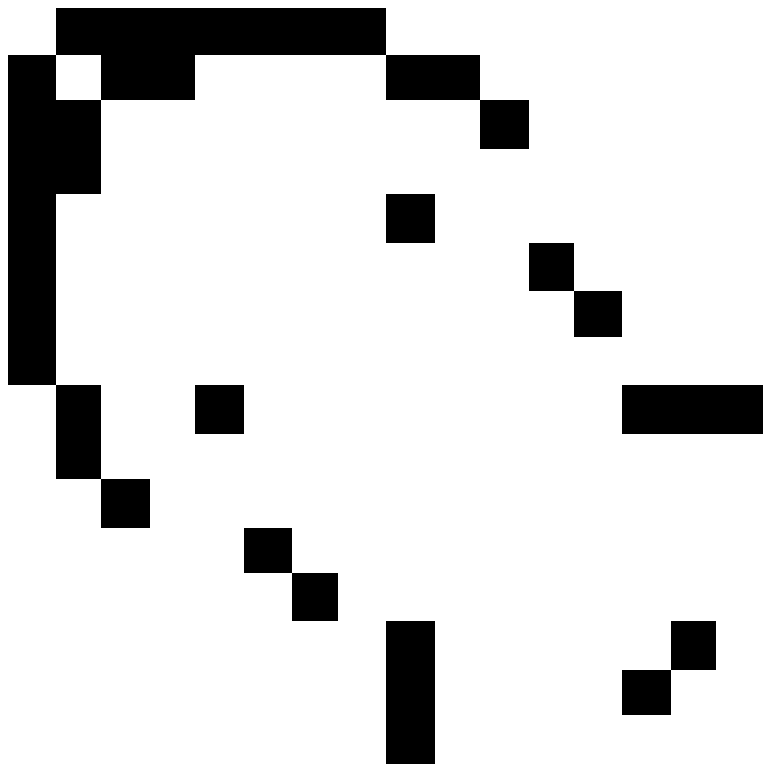}}\\
\end{tabular}}
\caption{{\bf Network Signatures.}
Leftmost are the ``network signatures'' (as pictures) of Facebook
and Wikipedia at the 64-node and 16-node scales.
Moving from left to right we show how that signature evolves as the
network is perturbed with increasing number of edge-swaps \math{\delta}, from
10\% to \math{\infty}.
For details on how these signatures (network pictures) are created we refer
to \cite{wu2016network,hegdesig2018}.
Rightmost are the network signatures of random graphs with the same
vertex-degrees.
At the 64-node scale, there is a clear change in
signature from \math{N_0} to \math{N_\infty}, indicating that the
``coordinated'' structure has been disrupted.
For Facebook, more so than Wikipedia, there is a significant change to
the signature even for just 10\% edge-swaps, which suggests Facebook
is more ``fragile'' at this scale.
At the 16-node scale, the signatures don't significantly change with
increasing edge-swaps,
suggesting that the structure at this scale is not coordinated enough
to be disrupted by random edge-swaps.
\label{fig:face-wiki}}
\end{figure}

The hard task is in Step 3, which poses a graph classification problem.
Any classifier trained in
Step 3 gives an estimate
\math{\hat\Delta(\kappa,\delta)\le\Delta(\kappa,\delta)}.
In \cite{wu2016network,hegdesig2018}, a variety of approaches to graph
classification are tested ranging from logistic regression and
random forests using classical graph features (average degree,
clustering coefficient, assortativity, etc.), to
graph kernels, to deep convolutional networks (CNN) using
lossless image representations of graphs 
proposed in~\cite{wu2016network,hegdesig2018}.\footnote{%
In a nutshell, graph images are formed from 
the \math{\kappa\times\kappa} adjacency matrix of a \math{\kappa}-node
subgraph (1's are black pixels and 0s are white
pixels). To structure the image into a signature which is invariant to
isomorphism, one must order the vertices canonically, and the ordering
which works best is  based on a modified BFS with preference to
high-degree nodes, see \cite{wu2016network,hegdesig2018} for details.}
The best performing method
is the CNN using the graph-image feature
from \cite{wu2016network,hegdesig2018}, and a close second is
logistic regression on classical features.
Choosing features is not easy, and can depend on the
graph domain, hence we use the image representation in
\cite{wu2016network,hegdesig2018} which is general and lossless. The CNN
extracts appropriate features from this powerful graph image and
learns a classifier.
Using these graph images, Figure~\ref{fig:face-wiki} illustrates
how structure is perturbed
with increasing edge-swaps for subgraphs from Facebook
(a tightly structured network) and Wikipedia
(a loosely structured network).

We make some qualitative observations
from the pictures in Figure~\ref{fig:face-wiki}.
Networks have signatures at different scales.
As one perturbs a network, signature changes 
are  visually
discernible.
Thus, a powerful  CNN classifier using these graph
image-signatures should come close 
to optimal classification accuracy. 
Further,
different networks have different levels of structure at different
scales. For example the Facebook signature at the 64-node scale is
significantly disrupted by 10\% edge-swaps, while the
Wikipedia signature is not as disrupted. The level to which
the signature at scale \math{\kappa}
gets disrupted by \math{\delta} edge-swaps is captured by
\math{\Delta(\kappa,\delta)}, so we expect
\mandc{
\Delta_{\text{Facebook}}(64,10\%)\gg \Delta_{\text{Wikipedia}}(64,10\%).
}
At a small enough scale, the signature does not
significantly change (e.g.
the 16-node signatures in Figure~\ref{fig:face-wiki}).
This suggests there is
a critical scale \math{\kappa^*} at which the signature change becomes
discernible with high accuracy.


Our experimental design is quite simple.
For each network and for each pair of values 
\math{(\kappa,\delta)}, where
\mandc{
\begin{array}{rcl}
\kappa&\in&\{4,5,6,\ldots,64\}\\
\delta&\in&\{10\%, 20\%, 30\%, 40\%, 50\%, \infty\},
\end{array}
}
we estimate \math{\Delta(\kappa,\delta)} using
\math{\hat\Delta(\kappa,\delta)} from Algorithm~1.
Note that \math{\delta} is a percentage of the number of
edges in the network, allowing us to compare networks of different
sizes.
We repeat each experiment
for each network 10 times to reduce the variance due randomness in
the construction of \math{N_\delta} and the sampling of
subgraphs to create training and test sets.
In all cases, the learning algorithm is the CNN using the
graph-image features, as already described earlier. For comparison, we
also show some results for classifying based on topological graph features
such as clustering coefficient and assortativity.

\remove{

\begin{figure*}[ht!]
  \centering
\begin{subfigure}{.125\textwidth}
  \centering
  \fboxsep0pt\fbox{\includegraphics[scale=0.25]{images/facebook_0}}
  \caption{$\delta=0\%$}
  \label{face0}
\end{subfigure}%
\begin{subfigure}{.125\textwidth}
  \centering
  \fboxsep0pt\fbox{\includegraphics[scale=0.25]{images/facebook_10}}
  \caption{$\delta=10\%$}
  \label{face10}
\end{subfigure}%
\begin{subfigure}{.125\textwidth}
  \centering
  \fboxsep0pt\fbox{\includegraphics[scale=0.25]{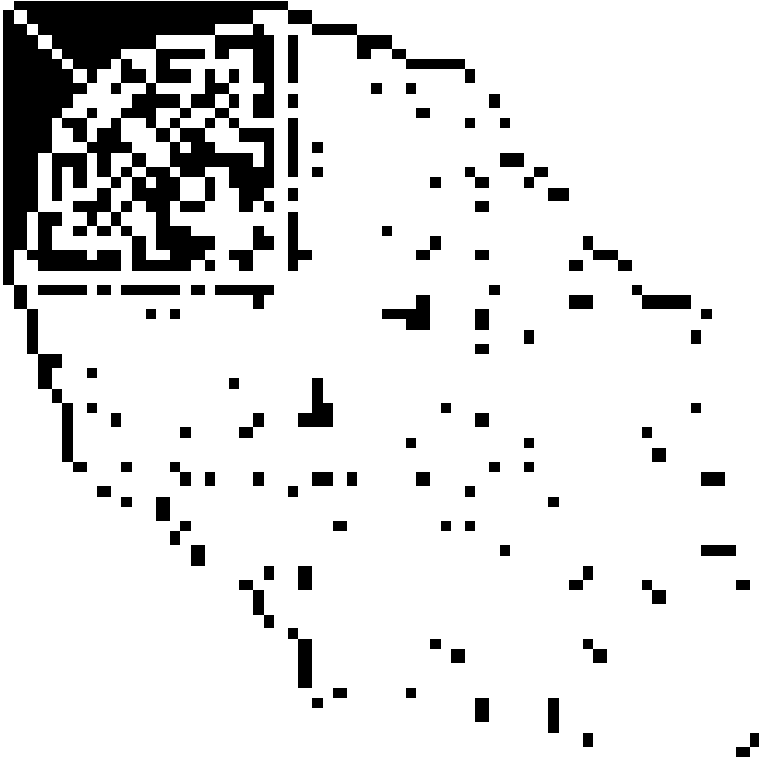}}
  \caption{$\delta=20\%$}
  \label{face20}
\end{subfigure}%
\begin{subfigure}{.125\textwidth}
  \centering
  \fboxsep0pt\fbox{\includegraphics[scale=0.25]{images/facebook_30}}
  \caption{$\delta=30\%$}
  \label{face30}
\end{subfigure}%
\begin{subfigure}{.125\textwidth}
  \centering
  \fboxsep0pt\fbox{\includegraphics[scale=0.25]{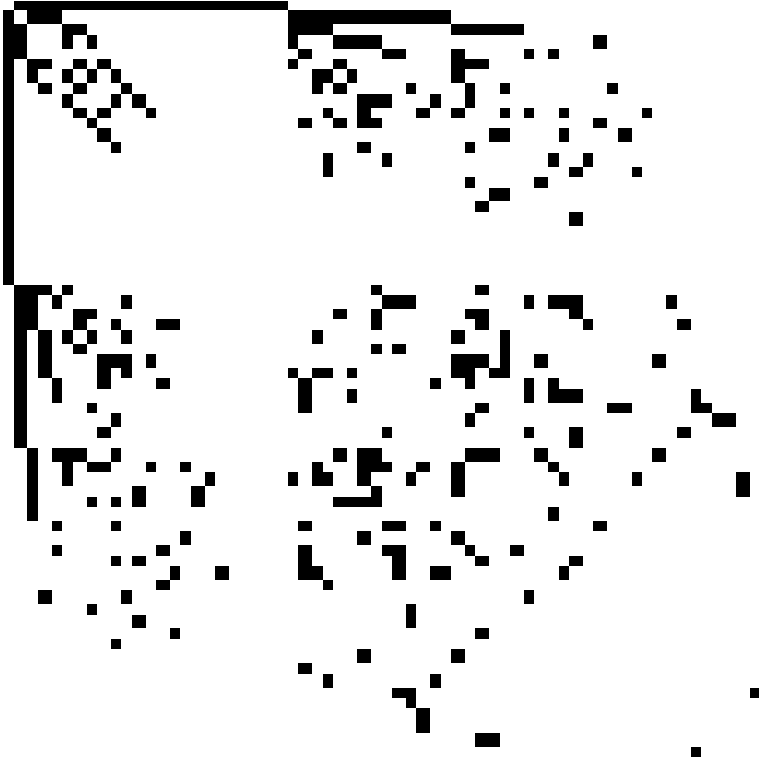}}
  \caption{$\delta=40\%$}
  \label{face40}
\end{subfigure}%
\begin{subfigure}{.125\textwidth}
  \centering
  \fboxsep0pt\fbox{\includegraphics[scale=0.25]{images/facebook_50}}
  \caption{$\delta=50\%$}
  \label{face50}
\end{subfigure}%
\begin{subfigure}{.125\textwidth}
  \centering
  \fboxsep0pt\fbox{\includegraphics[scale=0.25]{images/facebook_random}}
  \caption{$\delta=\infty$}
  \label{faceran}
\end{subfigure}%
\begin{subfigure}{.125\textwidth}
  \centering
  \fboxsep0pt\fbox{\includegraphics[scale=0.25]{images/eigen_facebook_64}}
  \caption{Top PC}
  \label{eigface64}
\end{subfigure}

\begin{subfigure}{.125\textwidth}
  \centering
  \fboxsep-2pt\fbox{\includegraphics[scale=0.23]{images/facebook_16_0}}
  \caption{$\delta=0\%$}
  \label{face160}
\end{subfigure}%
\begin{subfigure}{.125\textwidth}
  \centering
  \fboxsep-2pt\fbox{\includegraphics[scale=0.23]{images/facebook_16_inf}}
  \caption{$\delta=10\%$}
  \label{face1610}
\end{subfigure}%
\begin{subfigure}{.125\textwidth}
  \centering
  \fboxsep-2pt\fbox{\includegraphics[scale=0.23]{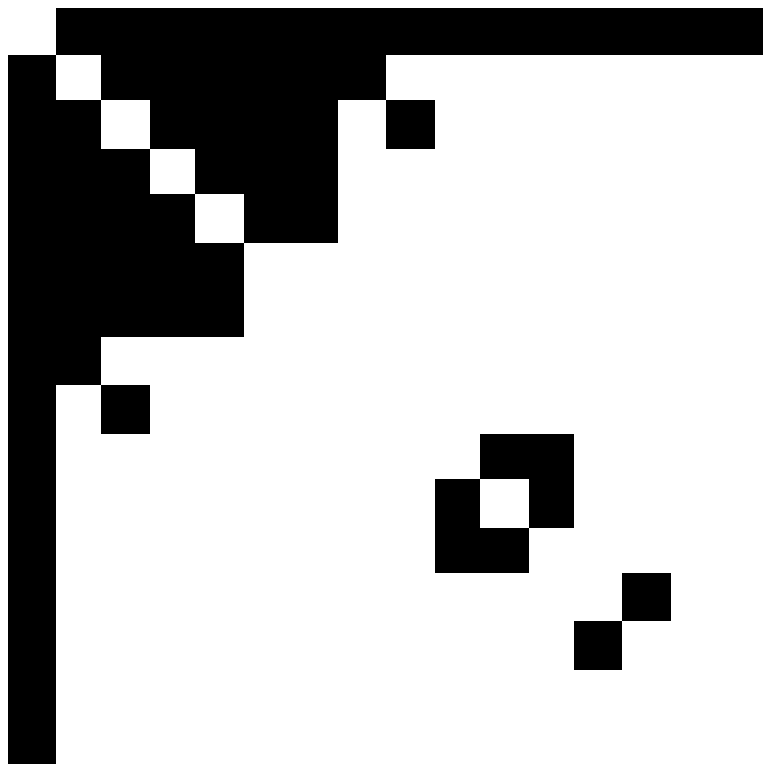}}
  \caption{$\delta=20\%$}
  \label{face1620}
\end{subfigure}%
\begin{subfigure}{.125\textwidth}
  \centering
  \fboxsep-2pt\fbox{\includegraphics[scale=0.23]{images/facebook_16_30}}
  \caption{$\delta=30\%$}
  \label{face1630}
\end{subfigure}%
\begin{subfigure}{.125\textwidth}
  \centering
  \fboxsep-2pt\fbox{\includegraphics[scale=0.23]{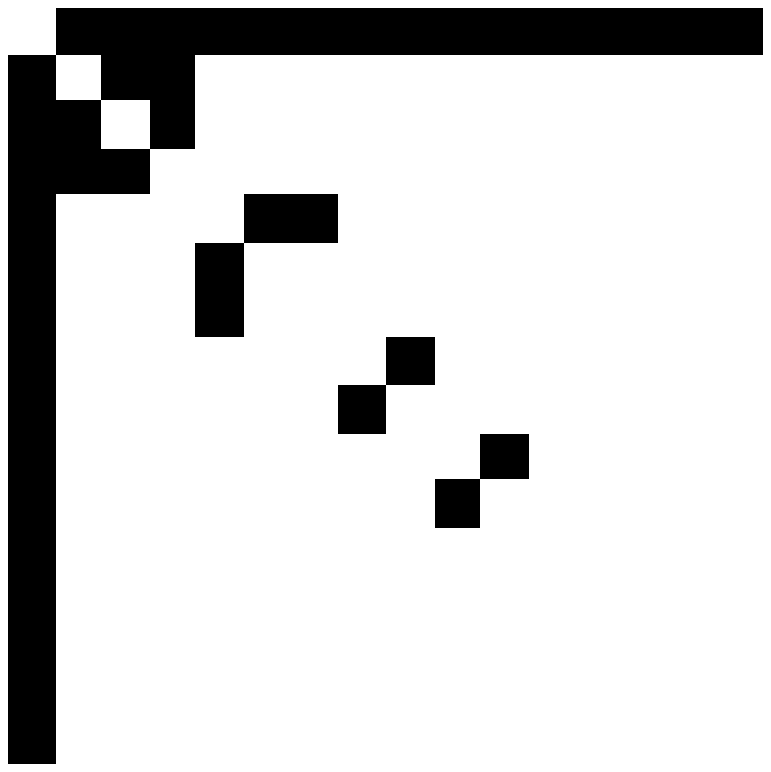}}
  \caption{$\delta=40\%$}
  \label{face1640}
\end{subfigure}%
\begin{subfigure}{.125\textwidth}
  \centering
  \fboxsep-2pt\fbox{\includegraphics[scale=0.23]{images/facebook_16_50}}
  \caption{$\delta=50\%$}
  \label{face1650}
\end{subfigure}%
\begin{subfigure}{.125\textwidth}
  \centering
  \fboxsep-2pt\fbox{\includegraphics[scale=0.23]{images/facebook_16_10}}
  \caption{$\delta=\infty$}
  \label{face16ran}
\end{subfigure}%
\begin{subfigure}{.125\textwidth}
  \centering
  \fboxsep0pt\fbox{\includegraphics[scale=0.25]{images/eigen_facebook_16}}
  \caption{Top PC}
  \label{eigface16}
\end{subfigure}
\caption{Progression of the Facebook network from original ($\delta=0\%$) to random ($\delta=\infty$) for $\kappa=64$ (top row) and $\kappa=16$ (bottom row). It is clear to see that most of the cliques (denoted by the dense black region in the top left corner of the images \ref{face0} - \ref{face20}) is gone before $\delta$ is raised to 30\%. This key difference which is clear to the human eye is picked up by the CNN which results in $\tau$ jumping from $\sim0.75$ at $\delta=10\%$ to $\sim0.9$ for $\delta=30\%$ for $\kappa=16$ (See Figure \ref{faceresult}). Since these differences between $P$ and $\bar{P}$ are easy to see for higher $\kappa$, the result is higher $\tau$. We show the top principal component of the signature for $\kappa = 64, 16$ at the end of each row.}
\label{face}
\end{figure*}

\begin{figure*}[ht!]
  \centering
\begin{subfigure}{.125\textwidth}
  \centering
  \fboxsep0pt\fbox{\includegraphics[scale=0.25]{images/wikipedia_0}}
  \caption{$\delta=0\%$}
  \label{wiki0}
\end{subfigure}%
\begin{subfigure}{.125\textwidth}
  \centering
  \fboxsep0pt\fbox{\includegraphics[scale=0.25]{images/wikipedia_10}}
  \caption{$\delta=10\%$}
  \label{wiki10}
\end{subfigure}%
\begin{subfigure}{.125\textwidth}
  \centering
  \fboxsep0pt\fbox{\includegraphics[scale=0.25]{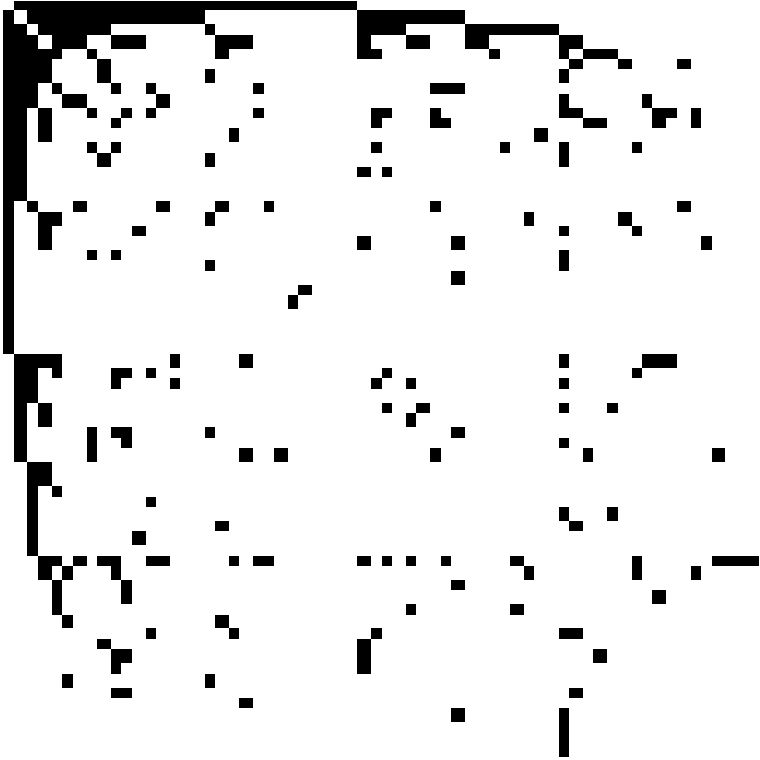}}
  \caption{$\delta=20\%$}
  \label{wiki20}
\end{subfigure}%
\begin{subfigure}{.125\textwidth}
  \centering
  \fboxsep0pt\fbox{\includegraphics[scale=0.25]{images/wikipedia_30}}
  \caption{$\delta=30\%$}
  \label{wiki30}
\end{subfigure}%
\begin{subfigure}{.125\textwidth}
  \centering
  \fboxsep0pt\fbox{\includegraphics[scale=0.25]{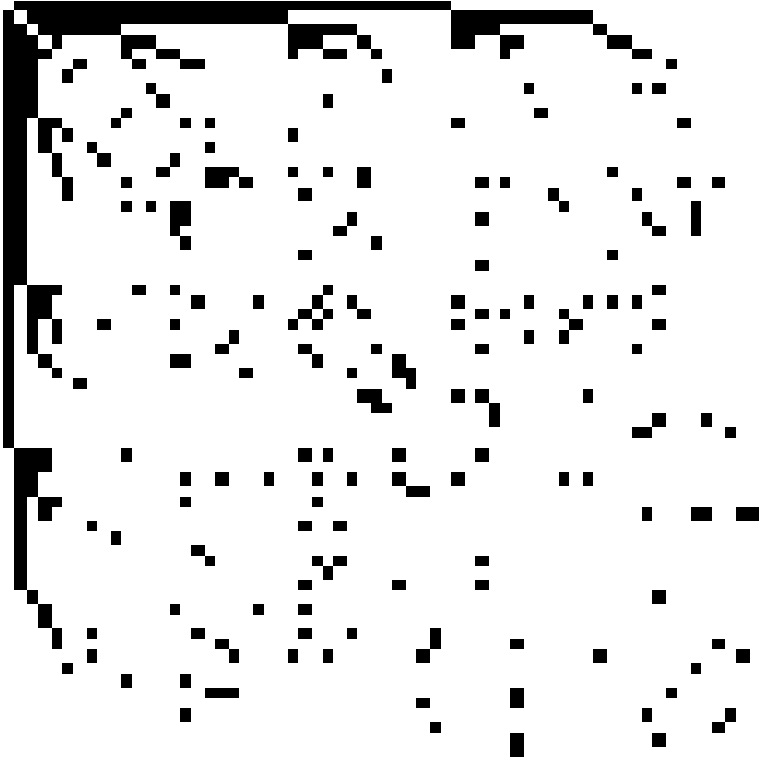}}
  \caption{$\delta=40\%$}
  \label{wiki40}
\end{subfigure}%
\begin{subfigure}{.125\textwidth}
  \centering
  \fboxsep0pt\fbox{\includegraphics[scale=0.25]{images/wikipedia_50}}
  \caption{$\delta=50\%$}
  \label{wiki50}
\end{subfigure}%
\begin{subfigure}{.125\textwidth}
  \centering
  \fboxsep0pt\fbox{\includegraphics[scale=0.25]{images/wikipedia_random}}
  \caption{$\delta=\infty$}
  \label{wikiran}
\end{subfigure}%
\begin{subfigure}{.125\textwidth}
  \centering
  \fboxsep0pt\fbox{\includegraphics[scale=0.25]{images/eigen_wiki_64}}
  \caption{Top PC}
  \label{eigwiki64}
\end{subfigure}

\begin{subfigure}{.125\textwidth}
  \centering
  \fboxsep-2pt\fbox{\includegraphics[scale=0.23]{images/wikipedia_16_0}}
  \caption{$\delta=0\%$}
  \label{wiki160}
\end{subfigure}%
\begin{subfigure}{.125\textwidth}
  \centering
  \fboxsep-2pt\fbox{\includegraphics[scale=0.23]{images/wikipedia_16_10}}
  \caption{$\delta=10\%$}
  \label{wiki1610}
\end{subfigure}%
\begin{subfigure}{.125\textwidth}
  \centering
  \fboxsep-2pt\fbox{\includegraphics[scale=0.23]{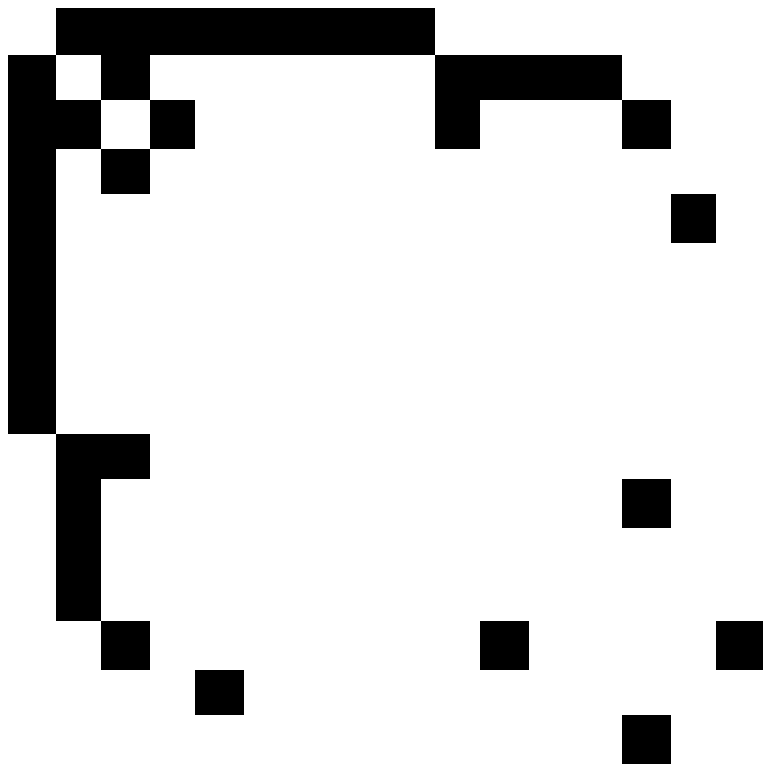}}
  \caption{$\delta=20\%$}
  \label{wiki1620}
\end{subfigure}%
\begin{subfigure}{.125\textwidth}
  \centering
  \fboxsep-2pt\fbox{\includegraphics[scale=0.23]{images/wikipedia_16_30}}
  \caption{$\delta=30\%$}
  \label{wiki1630}
\end{subfigure}%
\begin{subfigure}{.125\textwidth}
  \centering
  \fboxsep-2pt\fbox{\includegraphics[scale=0.23]{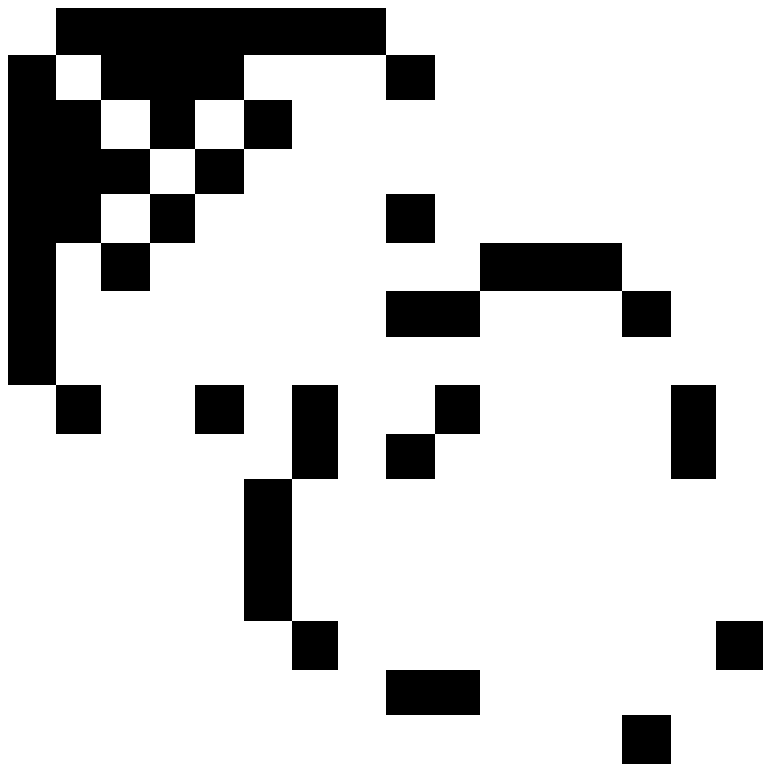}}
  \caption{$\delta=40\%$}
  \label{wiki1640}
\end{subfigure}%
\begin{subfigure}{.125\textwidth}
  \centering
  \fboxsep-2pt\fbox{\includegraphics[scale=0.23]{images/wikipedia_16_50}}
  \caption{$\delta=50\%$}
  \label{wiki1650}
\end{subfigure}%
\begin{subfigure}{.125\textwidth}
  \centering
  \fboxsep-2pt\fbox{\includegraphics[scale=0.23]{images/wikipedia_16_inf}}
  \caption{$\delta=\infty$}
  \label{wiki16ran}
\end{subfigure}%
\begin{subfigure}{.125\textwidth}
  \centering
  \fboxsep0pt\fbox{\includegraphics[scale=0.25]{images/eigen_wiki_16}}
  \caption{Top PC}
  \label{eigwiki16}
\end{subfigure}
  \caption{Progression of the Wikipedia network from original ($\delta=0\%$) to random ($\delta=\infty$) for $\kappa=64$ (top row) and $\kappa=16$ (bottom row). Since Wikipedia is not as structured as the Facebook network, it does not have any obvious features that pop out straight away. However, we see that raising $\delta$ has a very little effect on the signature of the Wikipedia network. That's why the human eye and the CNN model find it hard to distinguish between $P$ and $\bar{P}$ of Wikipedia. So, $\tau$ rises from just $\sim0.55$ at $\delta=10\%$ to $\sim0.6$ for $\delta=30\%$ for $\kappa=16$ (See Figure \ref{wikiresult}). Even for higher $\kappa$, $\tau$ is only about 0.65 which shows that Wikipedia has some inherent randomness about it. We show the top principal component of the signature for $\kappa = 64, 16$ at the end of each row.}
  \label{wiki}
\end{figure*}
}

\section*{Results and Discussion}
\label{results}

\begin{figure}[!h]
\centering
  \includegraphics[width=0.8\linewidth]{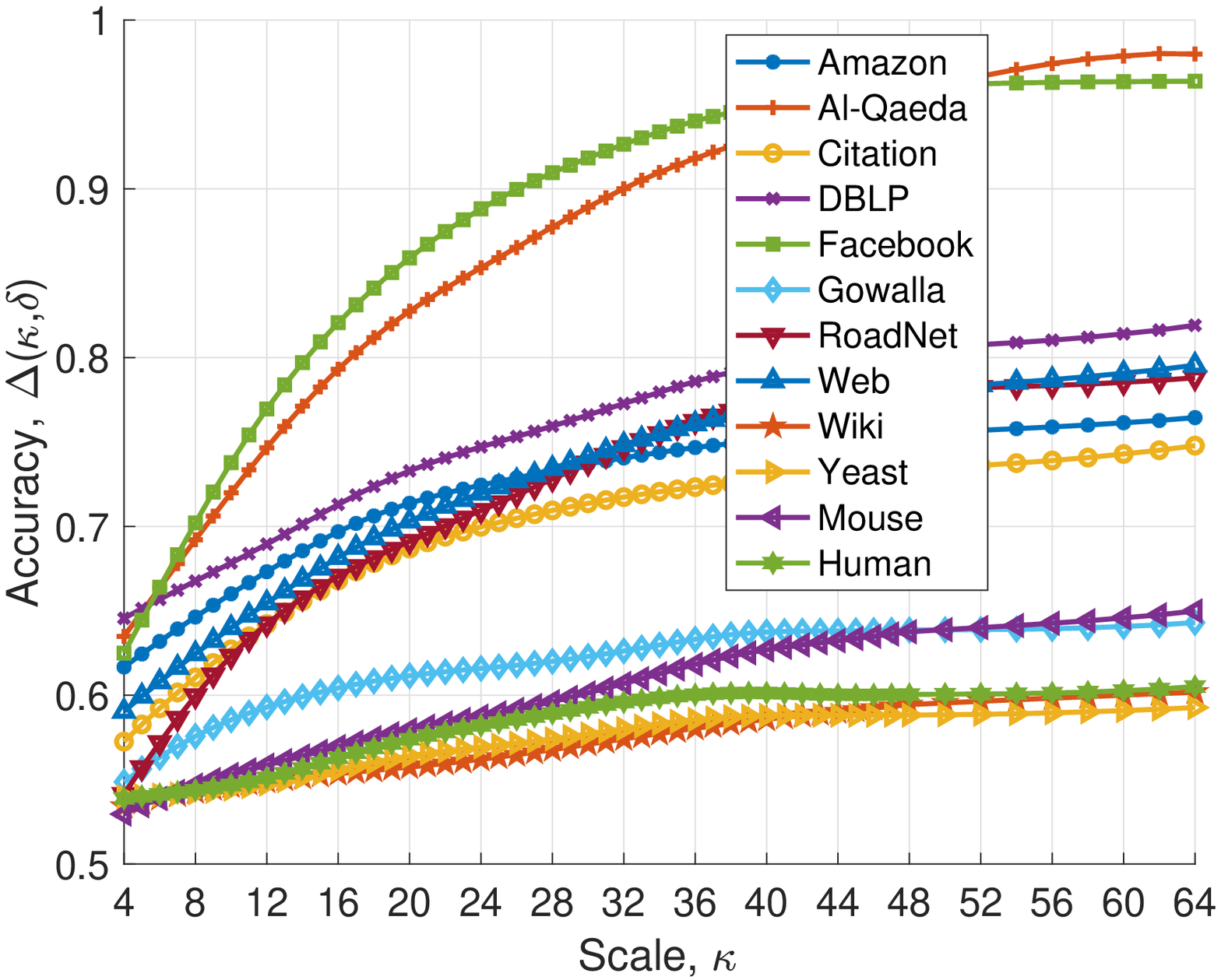}
  \caption{\label{fig:all-10}%
    The accuracy
    \math{\Delta(\kappa,10\%)} for all networks with \math{\delta=10\%}
    edge-swaps.
    We cluster the networks by the accuracy of distinguishing
    \math{N_0} from \math{N_\delta} at scale \math{\kappa=24}:
  }
    \fboxsep2pt\fbox{%
    \tabcolsep3pt%
  \begin{tabular}{lll}
  fragile&
  (\math{\Delta(24)>0.8})
  &
  Al-Qaeda; Facebook. 
  \\
  semi-robust&
  (\math{\Delta(24)\sim 0.7})
  &
  DBLP; Web; Road;\\
  & & Amazon; Citation.
  \\
  robust&
  (\math{\Delta(24)< 0.6})
  &
  Human; Yeast; Wiki;\\
  & & Mouse; Gowalla.
\end{tabular}}
\end{figure}

\paragraph{Robustness.}
We first show results for small perturbations.
Figure~\ref{fig:all-10} shows the
accuracy for just 10\% edge-swaps.

\begin{figure*}[t]
\centering
{\tabcolsep0pt\begin{tabular}{ccc}
  \includegraphics[width=0.33\textwidth]{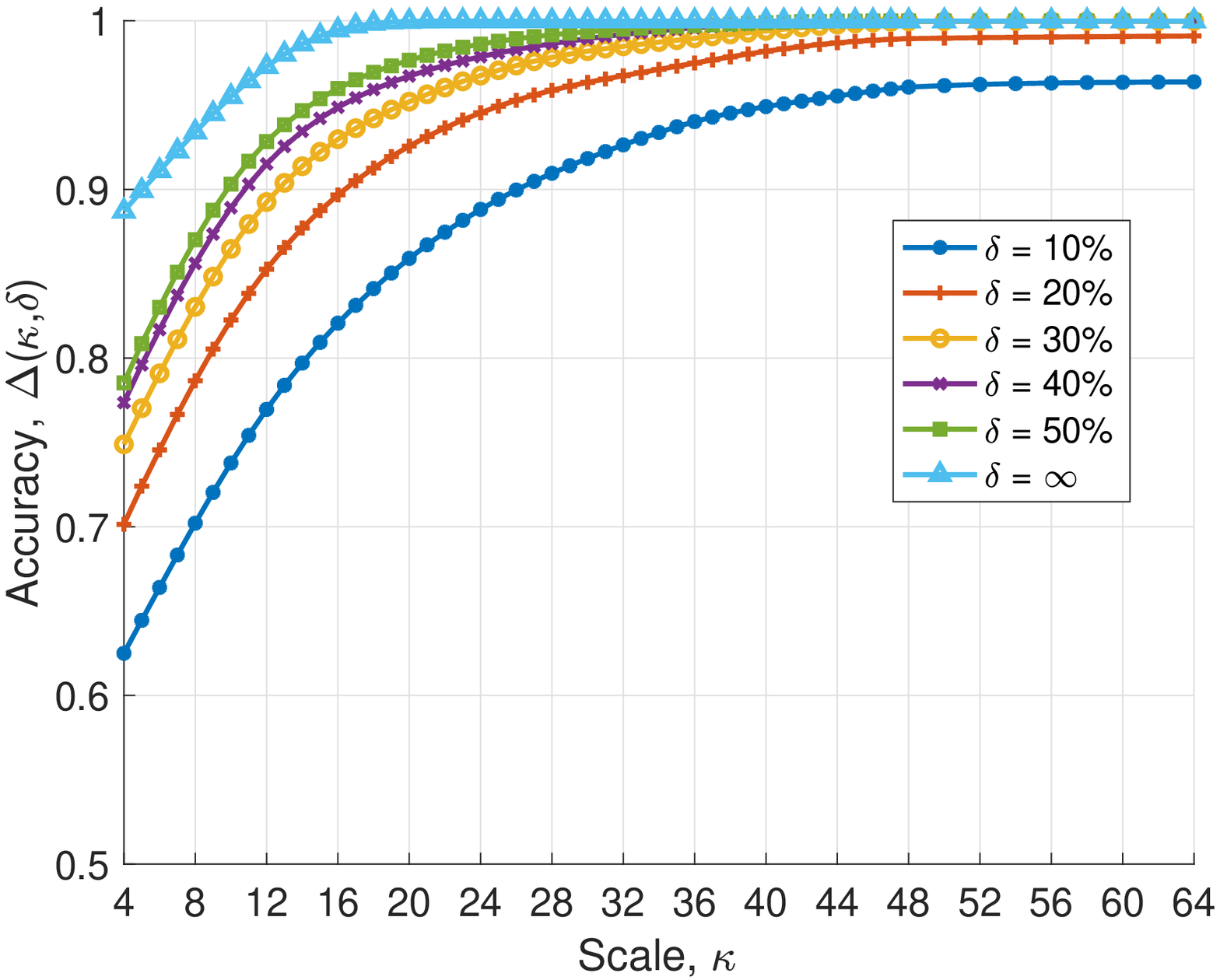}
  &
  \includegraphics[width=0.33\textwidth]{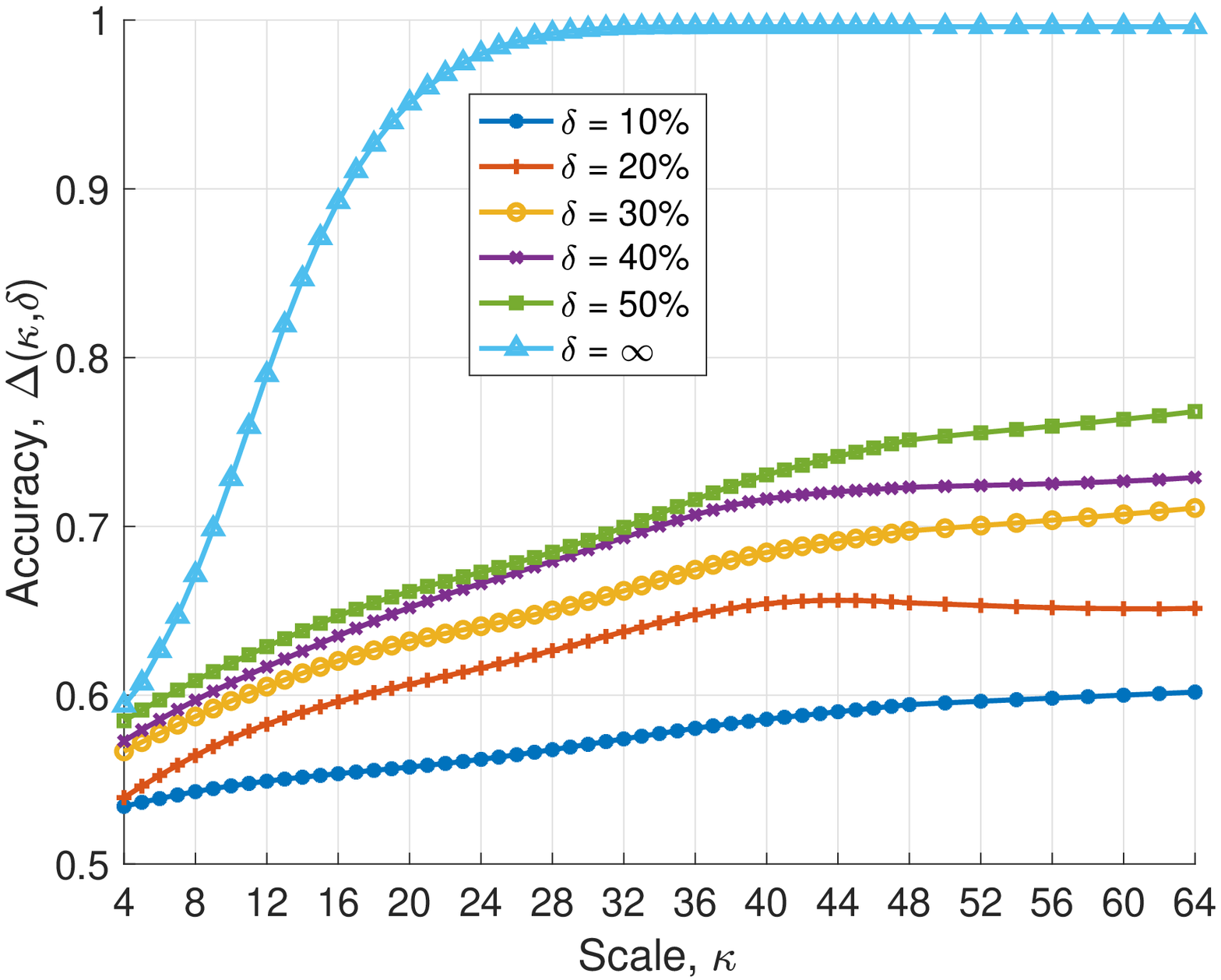}
  &
  \includegraphics[width=0.33\textwidth]{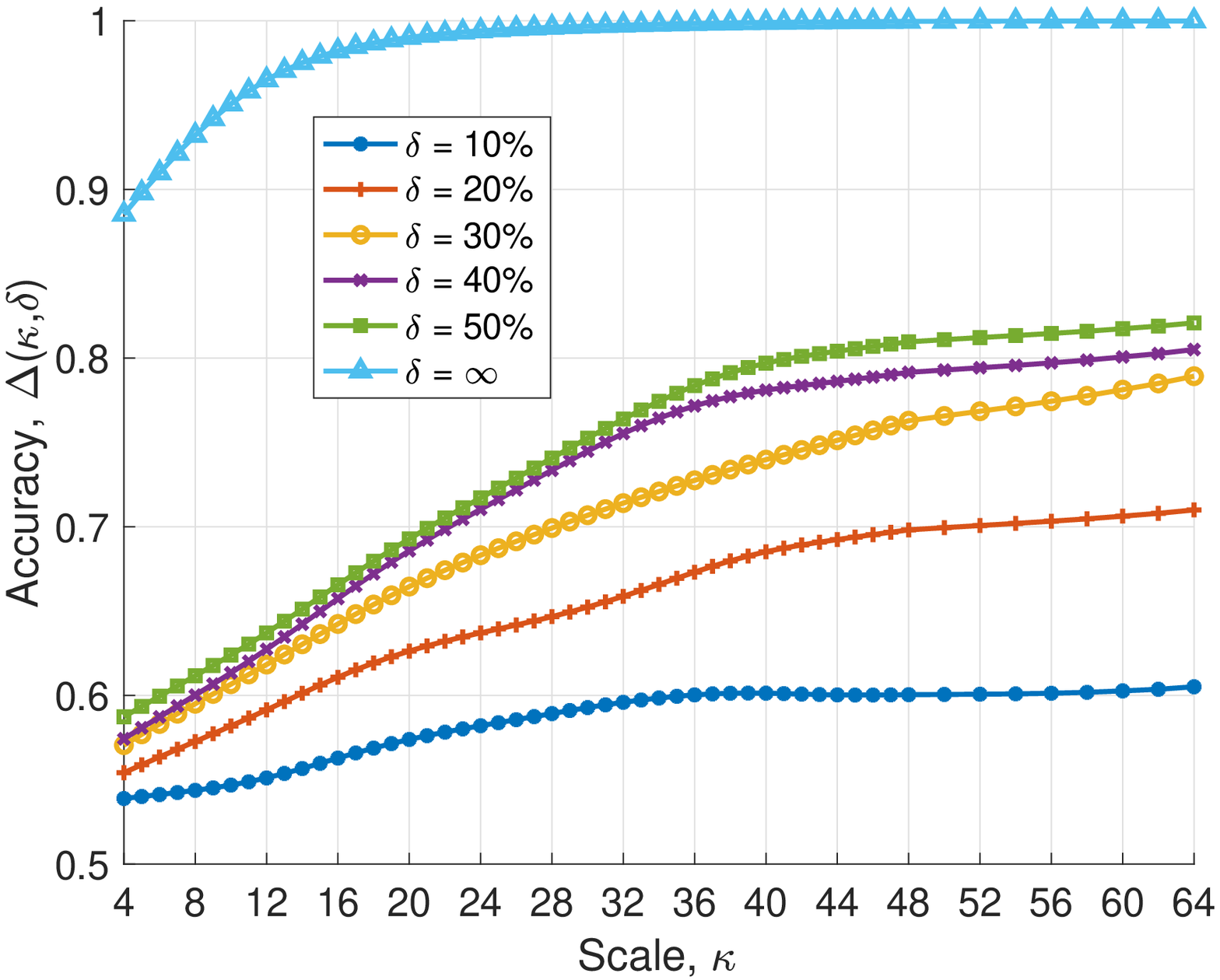}\\
  Facebook&Wiki&Human
\end{tabular}}
\caption{Evolution of a network's structure as it is perturbed from
  \math{\delta=10\%} to \math{\infty}.
  A fragile network (Facebook)
  with intricate structure quickly becomes distinguishable
  with perturbation, even at small scale. A robust network (Wiki) resists
  the perturbations and remains somewhat indistinguishable from the
  original network until \math{\delta} gets very large.}
\label{fig:delta-dep3}
\end{figure*}

\noindent
Recall that a high accuracy, \math{\Delta(\kappa)\gg\frac12}, means the
perturbed network at scale \math{\kappa} is highly distinguishable from the
original network. This means structure in the network has been disrupted.
Focusing on scale \math{\kappa=24} in Figure~\ref{fig:all-10},
we see that already at such a small scale,
\emph{for some networks}, there is significant
distinguishability between the original network \math{N}
and a 10\%-perturbed copy of \math{N}.
Indeed, the networks appear to cluster into
three groups which we categorize loosely as
fragile (\math{\Delta>0.8}), semi-robust
(\math{\Delta\sim 0.7}) and robust (\math{\Delta< 0.6}).
In the fragile networks, which are the social networks, a small perturbation
destroys the local structure leading to high
distinguishability. 
This may not be a surprise as people usually choose their
friends carefully and even small
perturbations will disrupt those finely tuned social circles --
this is especially so in the Al-Qaeda network
which achieves more than 90\% distinguishability with just 10\%
edge-swaps. In robust networks,
the distinguishability with just a 10\% perturbation is only marginally
above random. This does not mean there is no structure at the 24-node scale.
It just means the structure has not yet been significantly disrupted
by so small a perturbation. The biological networks fall into our
classification of robust, which may indicate a level of redundancy/degeneracy
that has been accumulated over the evolutionary process.
The semi-robust networks are also interesting
(DBLP, Web, Road, Amazon, Citation).
These networks do have structure, but that structure is not so fragile as the
social networks, indicating that the structure
is not as fine tuned. Indeed, these networks have grown in an ad-hoc
manner to represent the activity patterns of their actors, rather
than being explicitly created by their actors (cf. social networks).

Figure \ref{fig:delta-dep3} shows how structure gets dismantled
as the perturbation increases from \math{\delta=10\%} to \math{\delta=\infty}
for three networks:
Facebook (fragile social network); Wiki (robust information network);
and, a biological network. Facebook quickly yields and after
50\% edge-swaps the network has more-or-less reached a random graph with
the same degrees. The Wiki network, on the other hand, resists, and even after
50\% edge-swaps, the network is still not significantly discernible
from the original unperturbed network. The mixing time for the
edge-swapping random process is \emph{much slower} on the robust
Wiki network. The biological network resits small perturbations but slowly
yields its structure with larger perturbations.

\begin{figure*}[t]
\centering
{\tabcolsep0pt\begin{tabular}{ccc}
  \includegraphics[width=0.33\textwidth]{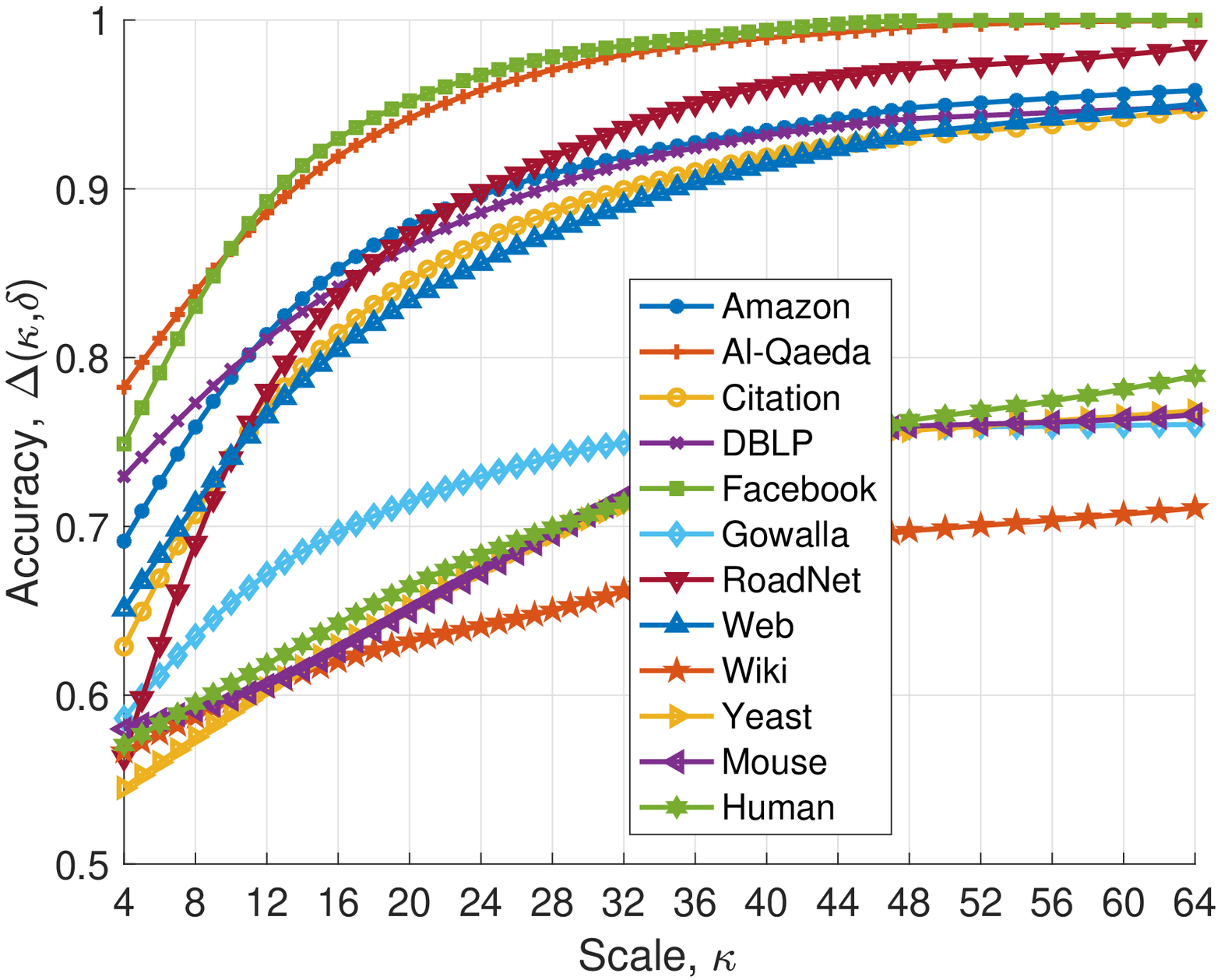}
  &
  \includegraphics[width=0.33\textwidth]{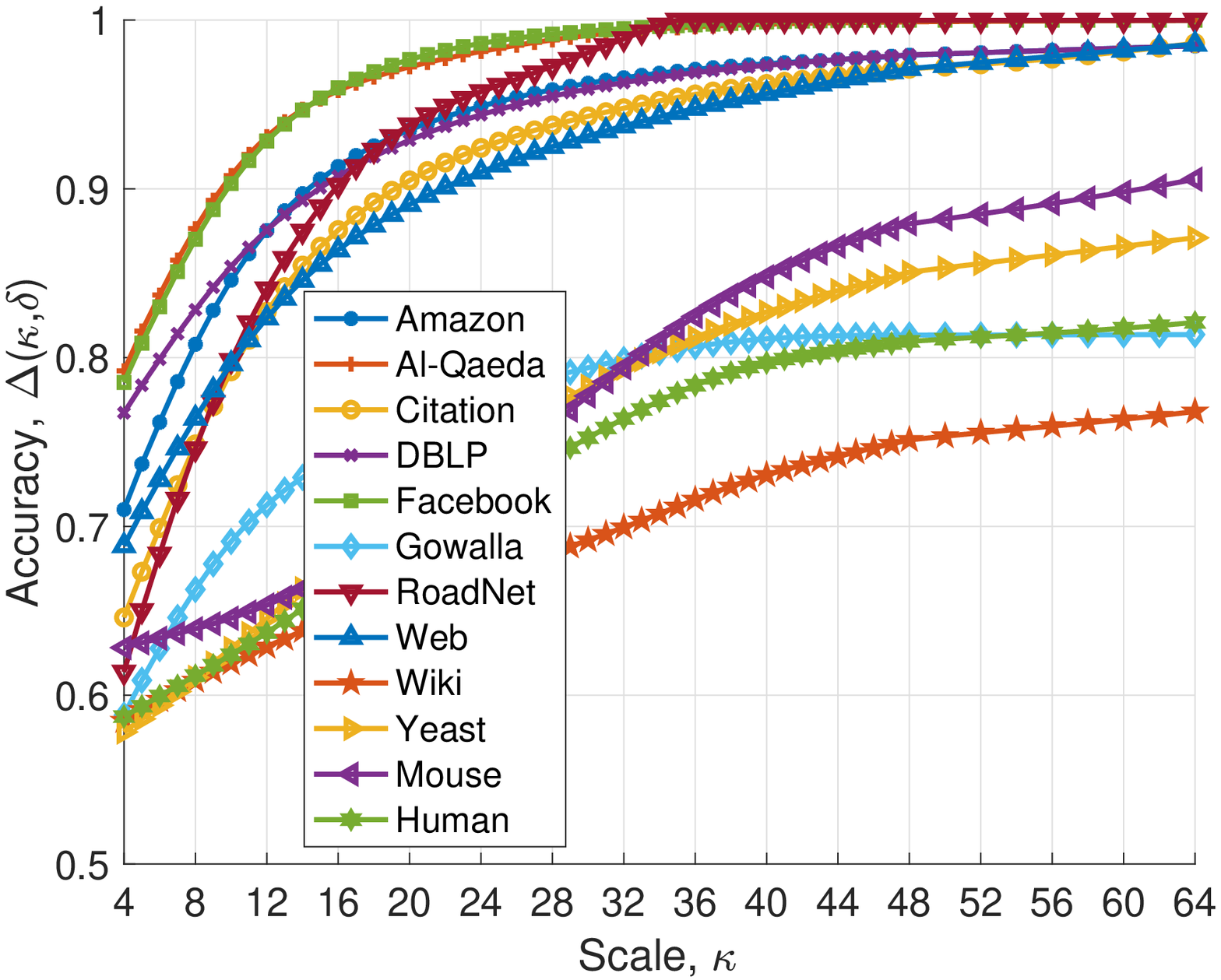}
  &
  \includegraphics[width=0.33\textwidth]{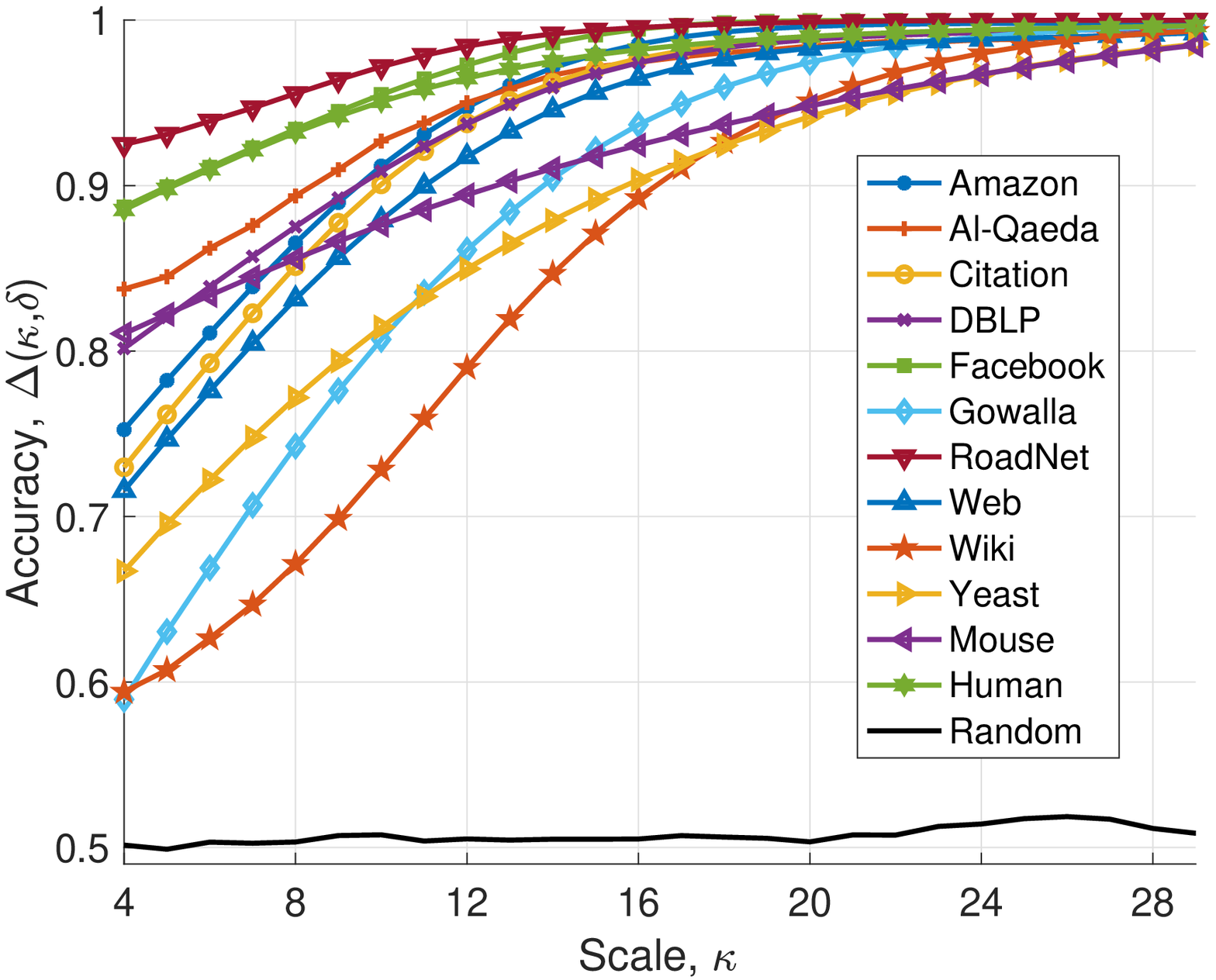}\\
  (a) \math{\delta=30\%}&(b) \math{\delta=50\%}&(c) \math{\delta=\infty}
\end{tabular}}%
\caption{Intrinsic scale at different levels of
  perturbation \math{\delta}.
  All networks display a ``rapid'' rise in accuracy
  (structure is rapidly lost) followed by an elbow followed by
  a flattening.
  As a sanity check, we also show the accuracy for an
  Erd\H{o}s-Renyi random graph at different scales for \math{\delta=\infty}.
  Such a graph has no ``structure'' at any scale,
  and it is no surprise that the
  accuracy hovers around \math{\frac12} for all scales.
\label{fig:all}}
\end{figure*}

\paragraph{Intrinsic Scale.}
The view presented in~Figure \ref{fig:delta-dep3}
highlights the evolution of a network as it is perturbed.
Some networks vigorously resist even at large scales (hard to
distinguish from the original network)
and some fall apart even at smaller scales (easy to distinguish from the
original network). We now go back to Figure~\ref{fig:all-10}
and focus on intrinsic
scale.
Figure~\ref{fig:all} shows results analogous to Figure~\ref{fig:all-10},
but for increasing values of the perturbation \math{\delta}.
The typical behavior is a rapid rise in accuracy as scale increases,
which corresponds
to a rapid dismantling of the networks structure. This is followed by an
elbow-turning point after which diminishing
returns results in a flattening.
The turning point (elbow) roughly corresponds to intrinsic
scale, the scale at which all the observable structure has been
dismantled by the perturbation  -- going to larger scale does not
improve accuracy significantly.

We now focus on \math{\delta=\infty} to define the intrinsic scale.
This choice of \math{\delta} is to capture all the structure, whether
fragile or robust -- we must
perturb hard enough to overcome the ``robustness'' of the network.
For small perturbations, inability to distinguish the
perturbed from the  non-perturbed subgraphs
may not indicate a lack of structure, but
just that whatever  structure exists may not yet have been dismantled.
At \math{\delta=\infty}, all existing structure beyond the vertex
degrees is gone. Indistinguishability now means there
was no structure to start with. Distinguishability with high accuracy
says that there was enough structure at the beginning. This structure
may have been fragile or robust, but at \math{\delta=\infty} we can't tell.

Visually looking at the elbows in Figure~\ref{fig:all} for
\math{\delta=\infty} suggests that the
networks roughly cluster into three categories.
\mandc{
\fboxsep2pt\fbox{%
\tabcolsep3pt%
\begin{tabular}{lp{0.5\linewidth}}
  Highly Structured
  &
  Road. (no surprise)
  \\
  Semi-structured&
  Al-Qaeda, DBLP; Web; Road; Amazon; Citation; Gowalla.
  \\
  Loosely structured
  &
   Wiki; Yeast; Mouse; Human.
\end{tabular}}}
Computing the elbow in the curves is not well defined and hard to
generalize, so we opt for a simpler definition of intrinsic scale:
the accuracy at \math{\delta=\infty} must be above 95\%.
This accuracy threshold is quite strict and an intrinsic scale
defined by the elbow will usually be smaller. Nevertheless, we opt
for this simpler and more conservative definition.
The intrinsic scales presented in~Table~\ref{tab:kappa} for
different accuracy thresholds can all
be obtained from  Figure~\ref{fig:all}.
The surprising conclusion is that for all these networks, spanning
a variety of domains, the intrinsic scale is no more than
20 and as low as 7.

It is also interesting to note from Figure~\ref{fig:all}(c) that the
accuracy
approaches but doesn't quite reach 1. This asymptotic gap away
from \math{1} indicates an amount of randomness in the original
graph that cannot be distinguished from the random graph. This gap
has about a 0.7 correlation with the intrinsic scale, and ranges from
\math{0.17\%} for the Road network to about \math{0.58\%} for the
Wiki network.
\remove{
\begin{table}
\centering
\begin{tabular}{c|c}
Network & \shortstack{(1 - max $\tau$)\\achieved for $\delta=\infty$}\tabularnewline
\hline 
Road & 0.0017\tabularnewline
Facebook & 0.0019\tabularnewline
Amazon & 0.002\tabularnewline
Terrorist Net. & 0.0021\tabularnewline
DBLP & 0.0027\tabularnewline
Yeast & 0.0028\tabularnewline
Web & 0.0033\tabularnewline
Citation & 0.0037\tabularnewline
Mouse & 0.0039\tabularnewline
Gowalla & 0.0042\tabularnewline
Wikipedia & 0.0058\tabularnewline
\end{tabular}
\caption{The accuracy gap (1 - max $\tau$ achieved for $\delta=\infty$}
\label{gap}
\end{table}
}

\begin{figure}[ht!]
\centering
{\tabcolsep-2pt\begin{tabular}{cc}
  \includegraphics[width=0.26\textwidth]{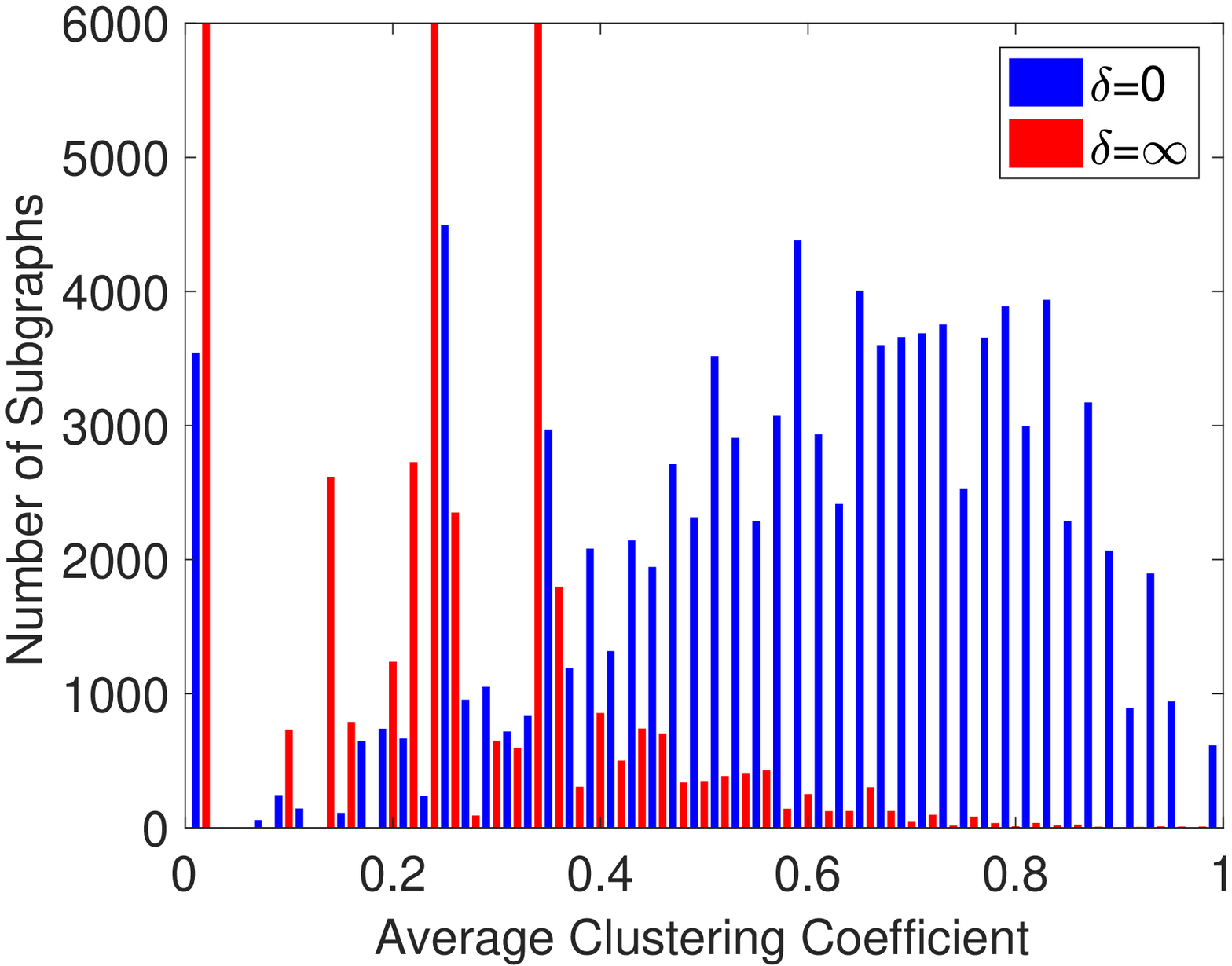}
  &
  \includegraphics[width=0.26\textwidth]{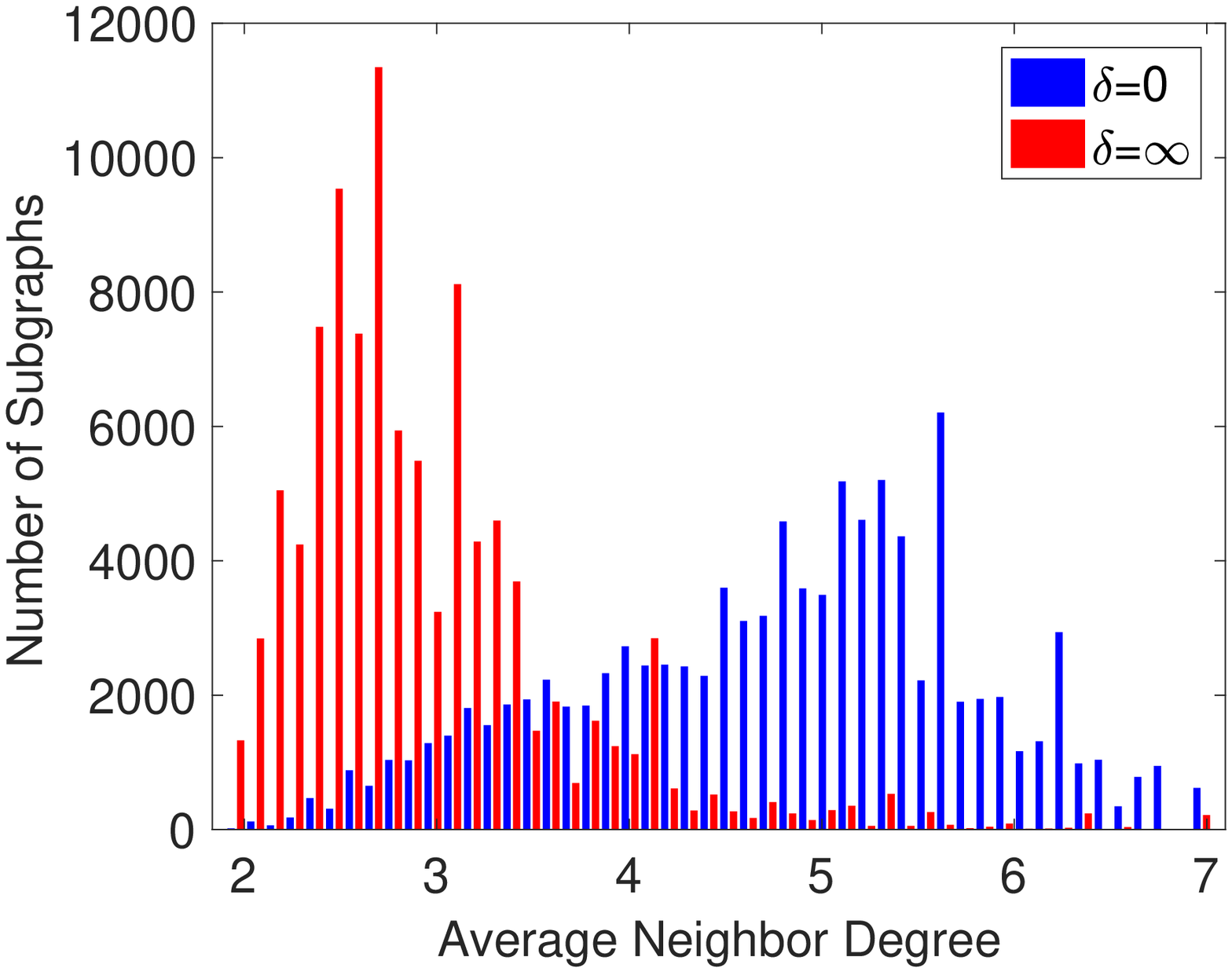}
\end{tabular}}
\caption{Classical features: histograms of clustering coefficient (fraction
  of closed triangles)
  and assortativity (average neighbor degree)
  for 8-node subgraphs of
  the Facebook network with \math{\delta=0} and \math{\delta=\infty}.
  The perturbed and non-perturbed subgraphs induce
  distinguishable distributions over these
features.}
\label{fig:classfea}
\end{figure}

\paragraph{Feature-Based Classification.}
Our algorithm to estimate \math{\Delta(\kappa,\delta)} uses
a learned classifier, and we have focused on the CNN with graph-images from
\cite{wu2016network,hegdesig2018}. We briefly compare with more traditional
feature-based methods. As a point of comparison, we take the
Facebook network with $\delta=\infty$, and consider two classical features:
\mandc{\text{\tabcolsep3pt
\begin{tabular}{lp{0.75\linewidth}}
  \emph{Clustering coefficient, \math{C}:}& Average fraction of
  closed\\
  & triangles per vertex.\\
  \emph{Measure of Assortativity, \math{r}:}& Average neighbor degree.
\end{tabular}}
}
We show histograms of these features for 8-node subgraphs of the
Facebook network and its perturbation
in Figure~\ref{fig:classfea}.

\noindent
The distributions \math{p_{8,0}} and \math{p_{8,\infty}} are clearly
distinguishable. We can compute the Bayes optimal accuracy for each feature
using \r{eq:bayes-optimal} where the sum over graphs \math{G} is replaced by
a sum of the feature's values. The results are in the table
below.
\mandc{
  \text{\tabcolsep3pt
    \begin{tabular}{l|l}
      Classifier&\math{\hat\Delta(8,\infty)}\\\hline
      Bayes optimal using \math{C}&0.905\\
      Bayes optimal using \math{r}&0.820\\
      Bayes optimal using \math{C} and \math{r}&0.932\\
      CNN \math{+} graph-image&\bf 0.934
    \end{tabular}
  }
}
The CNN with the graph-image gives the best (highest) estimate
\math{\hat\Delta}. 
Naturally, we can try other features and combinations of them,
but one cannot exhaust all the possibilities for any given network, and
further, a feature that works well for one type of network may not work well
for another.
And even still, there is no guarantee that the \emph{optimal} estimate
from using the features is better than the CNN plus graph-image.
The graph-image feature is general, lossless and
agnostic to the size and type of the network and when
combined with the CNN gives top performance.
Therefore CNN \math{+} graph-image was an 
easy choice for our classification problem.

\paragraph{Other Measures of Scale.}
Our intrinsic scale is not correlated with
network-size (the correlations are
negative: \math{-0.6}
with |V|, and
\math{-0.3} with |E|).
We compare our measure of intrinsic scale with other reasonable
measures of scale:
\mandc{\text{
\begin{tabular}{lp{0.55\linewidth}}
  \emph{Cluster size:}&
  Average of the  cluster-sizes from the Speakeasy algorithm in
  \cite{gaiteri2015identifying}.
  \\
  \emph{1-neighborhood size:}&Also the average degree,
  \math{2|E|/|V|}.
  \\
  \emph{Shortest path-length:}& Average over a large number of
  randomly sampled pairs of nodes.
  \\
  \emph{Network diameter:}& A measure of global scale.
\end{tabular}}}
We compare our intrinsic scale with these
measures below.\footnote{Average path length and diameter are estimated from a sample of 10\% of the vertex pairs.} 
\mandc{
  \text{\tabcolsep5pt
  \resizebox{0.5\textwidth}{!}{%
  \begin{tabular}{c||c|c|c|c|c}
Network& \shortstack{Intrinsic\\scale} & \shortstack{Cluster\\Size} & \shortstack{Neigh.\\Size}& \shortstack{Av. path\\length}& \shortstack{\\Diameter}\tabularnewline
\hline 
Road&7& 5.95 & 2.83 & 308.91 & 753\tabularnewline
Facebook&10 & 82.42 &43.7 & 3.83 & 7 \tabularnewline
Human&10 & 12.26 & 6.46 & 4.25 & 7\tabularnewline
Amazon&12 & 10.88 &5.53 & 11.97 & 31 \tabularnewline
Al-Qaeda&12& 8.47 &5.58 & 3.5 & 4 \tabularnewline
Cite&12& 48.45 & 24.4 & 4.36 & 10\tabularnewline
DBLP&13 & 9.77 &6.62 & 6.79 & 15\tabularnewline
Web&14 & 19.08 &11.7 & 6.34 & 16 \tabularnewline
Gowalla&17 & 17.65 &9.67  & 4.62 & 11\tabularnewline
Mouse&20 & 7.68 &2.86  & 4.86 & 10\tabularnewline
Yeast&20 & 18.56 & 16.9& 3.28 & 6\tabularnewline
Wiki&20& 199.65 &52.1 & 2.55 & 4\tabularnewline
\hline
corr. coef.&1.000 &0.3266  & 0.2257 & -0.5047 & -0.5043
\end{tabular}}}}
We also show the correlation coefficient of the other measures with intrinsic
  scale.
None of the other
measures are highly correlated with intrinsic scale. The closest is
cluster size which can be much larger and dependent on the clustering
algorithm.
Intrinsic scale captures something non-trivial.



\remove{
  \begin{figure*}[ht!]
\centering
%
\begin{subfigure}{.5\textwidth}
  \centering
  \includegraphics[width=\linewidth]{images/all-swaps50}
  \caption{Classification performance when $\delta=50\%$}
\end{subfigure}%
\begin{subfigure}{.5\textwidth}
  \centering
  \includegraphics[width=\linewidth]{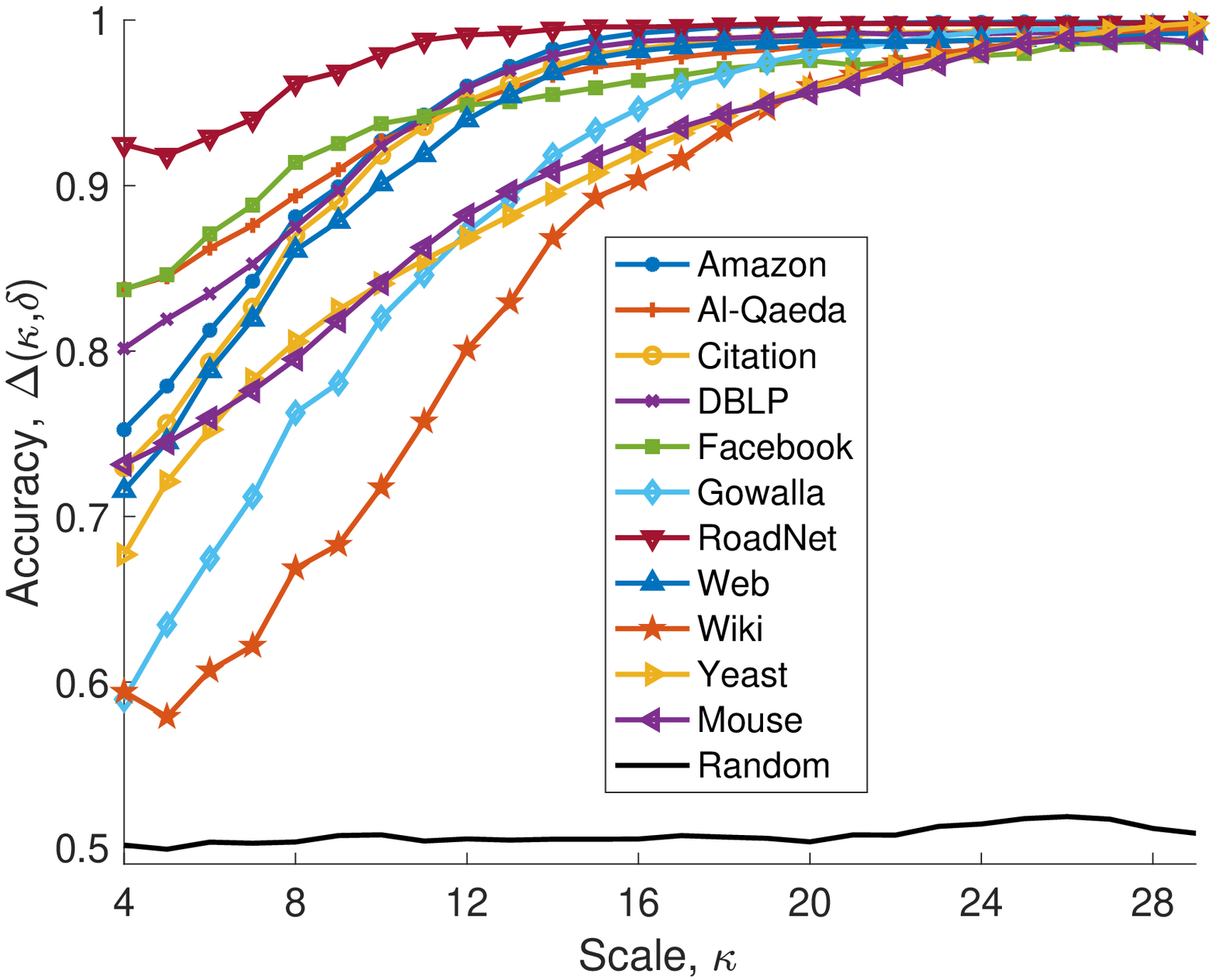}
  \caption{Classification performance when $\delta=\infty$}
\end{subfigure}
\caption{$\kappa^*(\delta=\{50\%, \infty\},\tau \in \{0.9, 0.95\})$ values in Table \ref{tab:kappa} are derived from these plots. We omit the plot for $\delta=30\%$ for brevity.}
\label{fig:results}
\end{figure*}
}


\section*{Conclusion}
\label{conclusions}

\begin{figure}[t]
  \centering
  \includegraphics[width=0.39\textwidth]{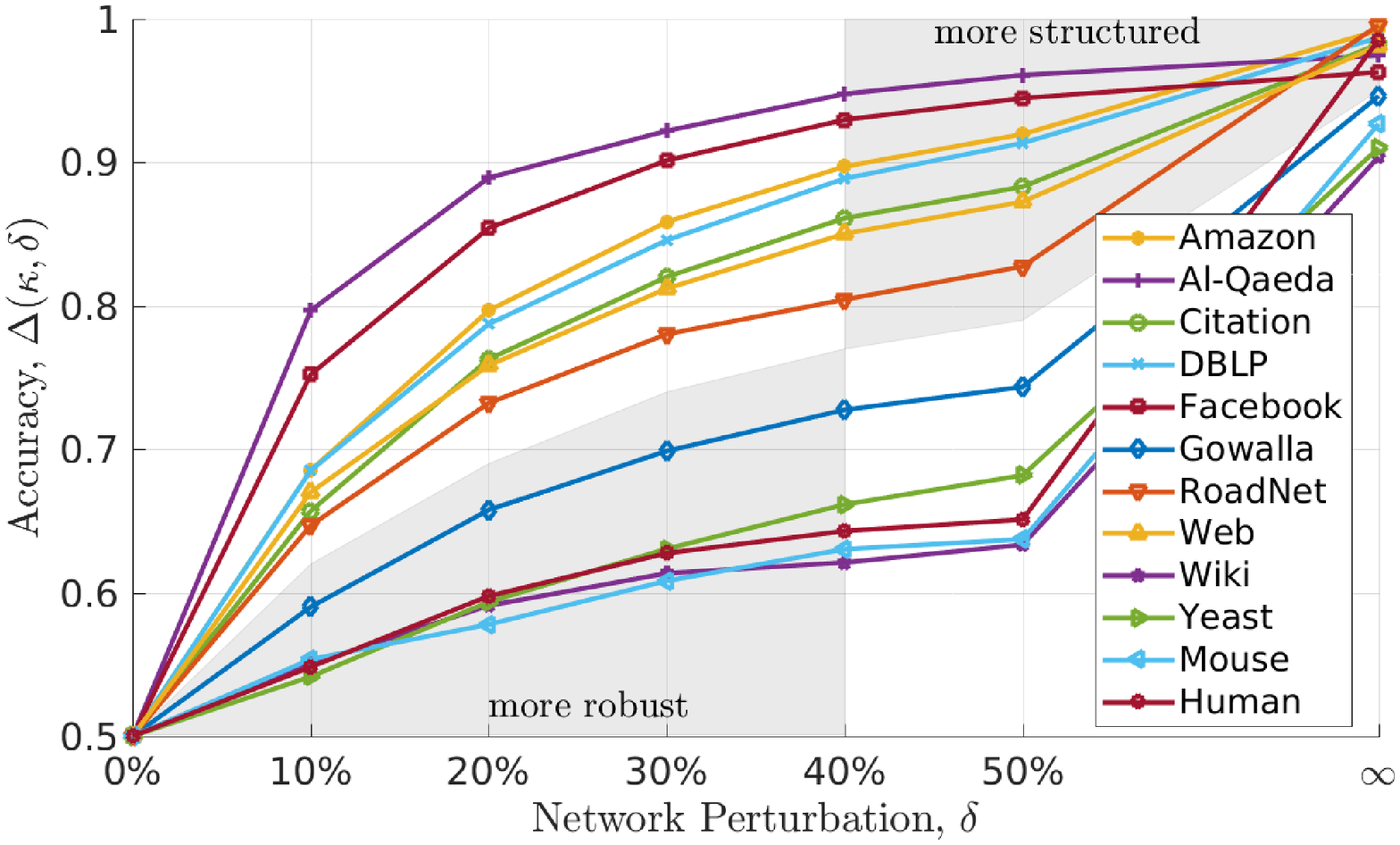}
  \caption{Learning curves for \math{\kappa=16}. Small perturbations
    reveal a network's robustness. Large perturbations,
    in particular \math{\delta=\infty}, reveal structure.
\label{fig:learning-curves}}    
\end{figure}

Our methodology for extracting the intrinsic scale of a network
poses the task as a classification problem.
This classification problem is to distinguish
subgraphs on the network from subgraphs on a perturbed copy
of the network.
The accuracy \math{\Delta(\kappa,\delta)} quantifies how much structure
in the network at scale \math{\kappa} gets dismantled by a
\math{\delta}-perturbation.
The learning curves for a fixed scale \math{\kappa}
in Figure~\ref{fig:learning-curves} show 
how the accuracy at that scale increases as one dismantles
the structure in the network (by increasing \math{\delta}).
The rate at which structure gets dismantled for small perturbations
is related to the robustness of the network, 
which we denote \math{\gamma}:
\mandc{
  \text{robustness, \math{\gamma(\kappa)}}=
  -\ln(\Delta(\kappa,10\%)-{\textstyle0.5}).
}
(logarithm(inverse of uplift in accuracy over random) for 10\%
perturbation). Robust networks hold on to their structure for small
perturbations. 
For large perturbations, all the structure gets
dismantled and the Bayes optimal accuracy quantifies the
amount of structure there was in the network to start with,
irrespective of robustness. We defined the intrinsic scale
\math{\kappa^*}
as the
scale at which there is enough structure
to achieve a classification accuracy exceeding 95\%.
A small intrinsic
scale means the network is very structured.

We summarize our
findings in the following graphic which represents the networks in our
study on 
a two-dimensional landscape of robustness and intrinsic scale.

\mandc{\includegraphics[width=0.55\linewidth]{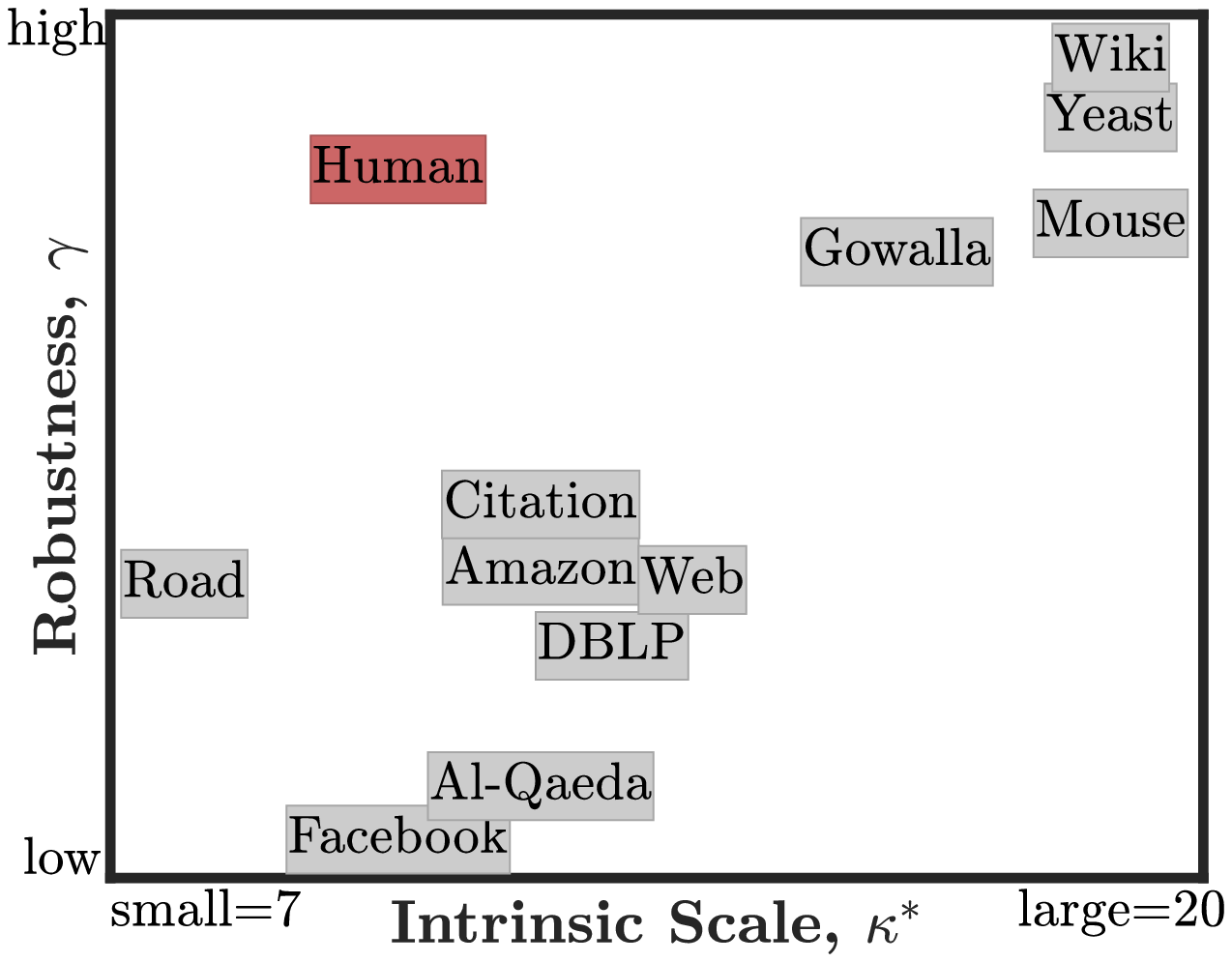}}

\noindent
The social networks are especially
fragile, and the biological networks are especially robust.
One can approximately quantify the resilience
of a network's functioning  to vertex and edge
removals using the degree-based parameter (see \cite{GBB2016}):
\mandc{\beta=\frac{\text{average squared degree}}{\text{average degree}}.}
There is a moderate correlation of 37\% between this measure of
resilience \math{\beta}
and our measure of robustness \math{\gamma}.
A correlation of 37\% indicates some relationship between a
network's ability to maintain its function under perturbation and
the statistical recognizability of a networks topology against a
null distribution obtained from a small (10\%) perturbation.
The relationship between statistical distinguishibility and
resilience may warrant further study (akin to the relationship between
statistical information and algorithmic compressability of sequences).

For the networks we examined, there is about a 61\%
correlation between structure and robustness. More structured networks
with smaller intrinsic scale tend to be less robust.
Our study provides a methodology for further investigation of 
this structure-robustness trade-off in networks.
The trade-off is by no means universal:
a notable exception is the Human PPI network which is very robust and yet
very structured. 

Interesting future directions are:
\begin{inparaenum}[(i)]
\item Using statistical distinguishability,
  one can construct a taxonomy of real
  networks and random models with respect
  to the structure-robustness trade-off. One might then
  identify which models are appropriate for different real networks.
\item How do we construct networks
  which break the structure-robustness trade-off,
  especially having very small intrinsic scale but very high robustness (e.g. Human PPI network). Such
  networks could have important applications.
\item
  One can use knowledge about the intrinsic scale of a network
  to inform other network
  analysis algorithms such as clustering.
  For example, clusters should be defined with respect to
  information available within the intrinsic scale of the nodes participating
  in the cluster. The intrinsic scale can also guide the choice of
  hyperparameters in clustering algorithms which set bounds for
  cluster sizes, etc. Since the intrinsic scales of real networks are small,
  algorithmic analysis of such networks, when confined to scales on the order
  of the intrinsic scale, should be more efficient.
\end{inparaenum}

\section*{Acknowledgment}
\label{sec:ack}
This research was supported by the Army Research Laboratory (ARL) under Cooperative Agreement W911NF-09-2-0053 (the ARL-NSCTA). The views and conclusions are those of the authors and do not represent the official policies, either expressed or implied, of ARL or the U.S. Government. The U.S. Government is authorized to distribute reprints for government purposes notwithstanding any copyright notation here on.

\bibliography{example_paper}
\bibliographystyle{icml2018}
\end{document}